\documentclass[preprint,aps,showpacs,floatfix]{revtex4}

\usepackage{epsfig}% Include figure files
\usepackage{dcolumn}% Align table columns on decimal point
\usepackage{bm}% bold math

\begin{document}

%\preprint{APS/123-QED}

\title{On the usefulness of the spectral function concept}

\author{
J.~Golak,
H.~Wita\l{}a,
R.~Skibi\'nski
}
\affiliation{M. Smoluchowski Institute of Physics, Jagiellonian University,
                    PL-30059 Krak\'ow, Poland}
\author{
W.~Gl\"ockle
}
\affiliation{Institut f\"ur Theoretische Physik II,
                 Ruhr Universit\"at Bochum, D-44780 Bochum, Germany}
\author{
A.~Nogga
}
\affiliation{Institute for Nuclear Theory, University of Washington, 
                  Box 351550 Seattle, WA 98195, USA}
\author{
H.~Kamada
}
\affiliation{Department of Physics, Faculty of Engineering,
Kyushu Institute of Technology,
1-1 Sensuicho, Tobata, Kitakyushu 804-8550, Japan}

\date{\today}

\begin{abstract}
The usefulness of the spectral function $S$ in the process 
${}^3{\rm He}(e,e'N)$
has been investigated in a kinematical regime constrained by 
the conditions that the three-nucleon (3N) center-of-mass energy
$E_{3N}^{c.m.} \le $ 150 MeV and the magnitude of the three-momentum 
transfer, $\mid \vec Q \mid  \le $ 600 MeV/c. 
Results based on a full treatment of the
final state interaction are compared to the spectral function
approximation. In the case of proton knockout in the direction of the photon
kinematical conditions have been identified where both response functions,
$R_L$ and $R_T$, can be well approximated  by $S$. These conditions
occur for  certain low missing momenta and missing energies but not in
all cases. So care is required. In case of neutron knockout only $R_T$ is
a candidate for an approximate treatment by $S$. In the case of $R_L$ the
concept of using $S$ is not valid in the studied kinematical regime. This
does not exclude the possibility that beyond that regime it
might be useful. Possible applications using $S$ for the extraction
of electromagnetic form factors of the nucleons are pointed out.
\end{abstract}

\pacs{21.45+v,21.10-k,25.10+s,25.20-x}

\maketitle

\section{\label{sec1}Introduction}

The $(e,e'N)$ reactions have  been widely analyzed in the past using
the concept of  the  spectral function. This quantity has been introduced
for instance in the work of \cite{diep75,diep76} in the context
of inclusive electron scattering on $^3$He. In the following it  has
been intensively investigated by C. Ciofi degli Atti and collaborators
\cite{ciof78,ciof80,ciofa80,ciof84,ciof95}
and P.U. Sauer and collaborators \cite{mei83,schu93,sau94} as well as other groups. 
For heavier systems there is  a rich literature
where that tool has  been also extensively used \cite{boffibook}. More recent
work can be found in \cite{kie97} and \cite{ciof02,ciofa02}. In \cite{kie97} effects of
polarizations are included. In no case the full final state interaction (FSI)
has been dealt with.

The concept of the spectral function  in $(e,e'N)$ reactions is based on
the simplifying assumption that the nucleon is knocked out as a free particle
and only the remaining nucleons  interact among themselves. Thus for a $^3$He
target only a final state interaction 
%among the neutron and one of the two protons taken into account.  
between two nucleons is considered.
Also the antisymmetrization of the
knocked out nucleon with the other two nucleons is neglected. This picture
appears to be reasonable if the  knocked out nucleon  receives all or essentially all
of the photon three-momentum, which moreover should be not too small. Of
course that simplification was also enforced in the past by the simple
fact that the complete final state interaction could not be controlled
numerically.

Integrating over a certain missing energy interval one defines
"momentum distributions". We put that quantity into quotes since it is
not the true momentum distribution inside for instance $^3$He. The reason is
the restricted integration interval even if the approximation
underlying the use of the spectral function would be justified.

Over the years it has become possible to take FSI 
among the three nucleons completely into account
in the case of $^3$He \cite{we-electron}.
We present such a solution  and critically investigate the simplified picture
leading to the spectral function. Our framework, however, is still non-relativistic,
which forces us to stay below the pion threshold, thus below about
150 MeV 3N c.m. energy. In order not to induce too high nucleon momenta,
which also would require a relativistic treatment, we  restricted the
three-momenta of the photon to the maximally allowed values of
600 MeV/c. Though this is already a too high value, we used it to get a first
indication whether there will be a tendency that at the higher momenta
the final state interaction might decrease. Also we expect
that this  violation will not be too severe to prevent a  reasonable
insight into the failure or validity of the  assumptions
underlying the simplistic picture of the  spectral function.

The paper is organized as follows. Section II is a brief reminder of
the definition of the spectral function and of the  complete
formulation  for the final state interaction in case of $^3$He.
The two relevant pairs of kinematical
variables  for $(e,e'N)$ processes are the missing momentum and missing energy,
$k$ and $E$, and the virtual photon momentum and its energy, $Q$ and $\omega$.  
So in Sec.~II we also illustrate  the
mappings of the two related regions in the $k-E$ and $Q-\omega$ planes.
In Sec.~III we compare the spectral function
under various kinematical conditions to results taking the full final state
interaction into account. This investigation is performed for proton
and neutron knockout from $^3$He. We summarize in Sec.~IV.

\section{\label{sec2}Theoretical Framework}

\begin{figure}[htb]
\begin{center}
\epsfig{file=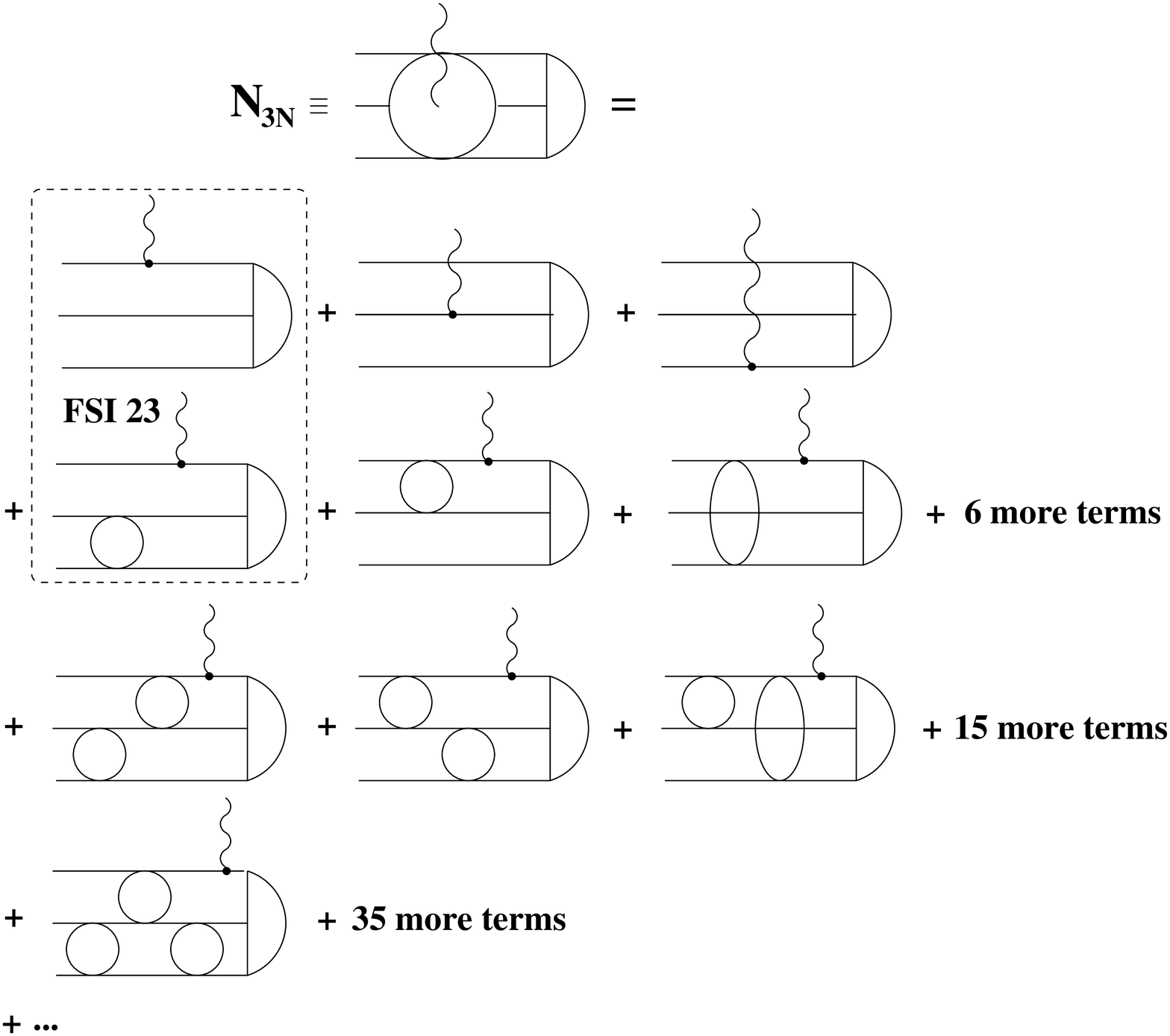,height=10cm}
\caption{\label{fig1}
Diagrammatic representation of the nuclear matrix element 
for the three-body electrodisintegration of $^3$He.
The open circles and ovals represent the two-body $t$-matrices.
Three horizontal lines between photon absorption and forces, and between
forces describe free propagation. The half-moon symbol on the very right
stands for $^3$He.
      }
\end{center}
\end{figure}

We regard the semi-exclusive process $^3{\rm He}(e,e'N)$ in parallel
kinematics, where the nucleon $N$ is knocked out with the momentum
${\vec p}_1$ parallel to the virtual photon momentum ${\vec Q}$.
In the unpolarized case the cross section is simply given as

\begin{eqnarray}
\frac{d ^6 \sigma}{ d E_{e'} d \Omega_{e'} d \Omega_{1}  d E_1 } =
 \sigma_{\rm Mott} \, \int d {\hat p} \, 
\left[  v_L R_L + v_T R_T \right] \, \frac{m^2 \, p \, p_1 }{2} ,
\label{eq1}
\end{eqnarray}
since the response functions $R_{TT}$ and $R_{TL}$ vanish under the
parallel condition~\cite{formula.xs1,formula.xs2}. 
The functions $v_L$ and $v_T$ are standard
kinematical factors.
The two response functions $R_L$ and $R_T$
are expressed in terms of the nuclear matrix elements $N_0$ and $N_{\pm 1}$ as

\begin{eqnarray}
R_L \equiv \frac12 \sum\limits_M \, \sum\limits_{m_1,m_2,m_3} 
\left| 
N_0 ({\vec p}_1, {\vec p}_2, {\vec p}_3 ; 
M, m_1,m_2,m_3; \nu_1,\nu_2, \nu_3 ) 
\right|^2 ,
\nonumber \\
R_T \equiv \frac12 \sum\limits_M \, \sum\limits_{m_1,m_2,m_3} 
\left(
\left| 
N_1 ({\vec p}_1, {\vec p}_2, {\vec p}_3 ; 
M, m_1,m_2,m_3; \nu_1,\nu_2, \nu_3 ) 
\right|^2  \right. 
\nonumber \\
+ \  \left.
\left| 
N_{-1} ({\vec p}_1, {\vec p}_2, {\vec p}_3 ; 
M, m_1,m_2,m_3; \nu_1,\nu_2, \nu_3 ) 
\right|^2
\right) ,
\label{eq2.1}
\end{eqnarray}
where $M$, $m_1$, $m_2$, $m_3$ are the initial $^3$He and final 3N 
spin magnetic quantum numbers, and $ \nu_1,\nu_2, \nu_3 $ are isospin magnetic
quantum numbers needed to identify the nucleons in the final state.
The direction (magnitude) of the relative momentum
of the two undetected nucleons
is denoted by ${\hat p}$ (p) and
the nucleon mass by $m$.
The matrix elements $N_0$ and $N_{\pm 1}$ 
are driven by the charge density operator and spherical components of the
transverse current operator, respectively. In general the nuclear
matrix element has the form
%with appropriate summation and averaging over spin directions
% and integration over the angles  p of the relative momentum
% of the two undetected nucleons. 
\begin{equation}
N^\mu \equiv
\langle \Psi_f^{(-)} \mid j^\mu ( \vec Q) \mid 
%\Psi^{\theta^* \phi^*}_{{}^3{\rm He}}
\Psi_{{}^3{\rm He}}
\rangle ,
\label{eq3}
\end{equation}
where $f$ comprises the momenta and the magnetic spin and isospin quantum numbers
of the three final nucleons. We shall concentrate here on 
the complete break up and refer the reader for the case of the
pd breakup to \cite{photon1}. As has been shown in \cite{photon1},
$N^\mu$ can be represented as

\begin{equation}
N^\mu = 
\langle  \phi_0 \mid ( 1 + P ) j^\mu ( \vec Q) \mid 
\Psi_{{}^3{\rm He}} \rangle  \, 
+ \,
\langle  \phi_0 \mid ( 1 + P ) \mid U^\mu \rangle ,
\label{eq4}
\end{equation}
where the auxiliary state $ \mid U^\mu \rangle$ obeys the Faddeev-like integral equation

\begin{eqnarray}
\mid U^\mu \rangle \, = \,
t G_0 (1 + P ) j^\mu ( \vec Q ) \mid  \Psi_{{}^3{\rm He}} \rangle  \, + \, 
t G_0 P \mid U^\mu \rangle .
\label{eq5}
\end{eqnarray}
The ingredients in Eq.~(\ref{eq5}) are the free 3N propagator $G_0$,
the NN $t$-operator generated via
the Lippmann-Schwinger equation from any modern NN interaction,
and a suitably chosen permutation operator $P$~\cite{gloecklebook}.
The state $\phi_0$ in Eq.~(\ref{eq4}) is a plane wave, antisymmetrized
in the two-body subsystem, where $t$ acts.
For a generalization including a three-nucleon force we refer
the reader to \cite{photon1}. In the present study we restrict ourselves
% to NN forces only but allow for one-and two-body currents  j  (Q).
to NN forces and allow only for one-body currents  $j_\mu ({\vec Q})$.
It is illustrative to present the physical content of the
expressions (\ref{eq4}) and (\ref{eq5}) in the following way. If one iterates
the integral equation and inserts the resulting terms into (\ref{eq4})
one arrives at the infinite sequence of processes shown in Fig.~\ref{fig1}.
In the first row there is no final state interaction and the
photon is absorbed by nucleons 1, 2 and 3. 
%For the sake of a simpler representation we did not add the diagrams with
%two-body currents.  
The next three rows include rescattering
processes of first order in the NN $t$-operator (denoted by a circle).
Then follow processes of second order in $t$, third order etc.
That complete sum of processes is generated by solving the
integral equation (\ref{eq5}). Now taking only the first diagrams in 
row 1 and 2 into account underlies the concept of the
spectral function $S$. The corresponding expression is

\begin{eqnarray}
N^\mu = \langle  \phi_0 \mid ( 1 + t_{23} G_0  ) j^\mu ({\vec Q};1) \mid 
\Psi_{{}^3{\rm He}}
\rangle ,
%\nonumber \\
%{\rm ???? \ 
%I \ do \ not \ know \ which \ formula \ would \ you \ like \ to \ have \ 
%at \ this \ point .     Dies ist okay und weiter unten fuehren wir dann die Definition
%von S ein}
%\nonumber \\
%S(k,E) = \frac{ m \, p}{2} \, \frac12 \sum\limits_M \, 
%\sum\limits_{m_1, m_2, m_3} \,
%\int d{\hat p} \,
%\left|
%\sqrt{6} \, \langle \nu_1 \nu_2 \nu_3 \mid 
%\langle m_1 m_2 m_3 \mid  \, \langle {\vec p} \, {\vec k} \mid
%( 1 + t_{23} G_0  ) \mid \Psi_{{}^3{\rm He}}
%\rangle 
%\right|^2
\label{eq6}
\end{eqnarray}
where the argument $1$ in the current explicitly indicates
that the photon is absorbed only on one nucleon, numbered 1
in our notation. That approximation, the two encircled diagrams
in Fig.~\ref{fig1}, will be called in the following FSI23 for short
and stands for final state interaction in the spectator pair $(23)$.
Related to that nuclear matrix element is the spectral function $S$. It is 
defined as

\begin{eqnarray} 
S(k,E) = \frac{ m \, p}{2} \, \frac12 \sum\limits_M \,
\sum\limits_{m_1, m_2, m_3} \,
\int d{\hat p} \,
\left|
\sqrt{6} \, \langle \nu_1 \nu_2 \nu_3 \mid
\langle m_1 m_2 m_3 \mid  \, \langle {\vec p} \, {\vec k} \mid
( 1 + t_{23} G_0  ) \mid \Psi_{{}^3{\rm He}}
\rangle
\right|^2
\label{eq6a}
\end{eqnarray}

%(????   Ist denn das Argument k im bra richtig?? Es steht doch vermutlich wegen des Strom
%operators zunaechst als Argumnet von 3He . Dann geht es durch $G_0$ und $t_{23}$ hindurch??
%Ja, dies ist wohl richtig , weil diese beiden Operatoren diagonal sind in der q-Variablen.
%Dies ist doch richtig?) 

The arguments of $S$ are the magnitude $k$ of the missing momentum 
\begin{equation}
k \equiv \mid {\vec Q} - {\vec p}_1 \mid
\label{eq7}
\end{equation}
and the excitation energy $E$ of the undetected pair. Nonrelativistically 
\begin{equation}
E \equiv \frac{p^2}{m} ,
\label{eq8}
\end{equation}
where $p$ is the relative momentum of the undetected nucleons.
Comparing the expression (\ref{eq6a}) for $S$
to the ones for $R_L$ and $R_T$ under the FSI23 approximation
%that only those two processes are taken into account 
one finds

\begin{eqnarray}
   S(k,E) = \frac12 \, m \, p \, \frac1{(G_E)^2} 
\int d {\hat p} R_L (FSI23) \nonumber \\
          = \frac12 \, m \, p \, \frac{2 m^2}{Q^2 (G_M)^2}  
\int d {\hat p} R_T (FSI23) .
\label{eq9}
\end{eqnarray}
This inserted into (\ref{eq1}) yields the well known relation between the cross
section and the spectral function
%form of the cross section under this simplifying assumption

\begin{eqnarray}
\frac{d ^6 \sigma}{ d E_{e'} d \Omega_{e'} d \Omega_{1} d d E_1 } =
\sigma_{\rm Mott} \, 
\left[  v_L (G_E)^2 + v_T \frac{Q^2 (G_M)^2}{2 m^2} \right] \, S (k, E) \, m \, p_1  \, \equiv \
\sigma_{eN} \, S (k, E) \, \rho_f .
\label{eq10}
\end{eqnarray}
Here the non-relativistic phase space factor $\rho_f$ is simply
\begin{eqnarray}
\rho_f = m \, p_1 \, \left( 1 + \frac{2 E_e}{m} \sin^2 \frac{\theta_e}{2} 
\right)
\label{eq11}
\end{eqnarray}
and the unpolarized electron-nucleon cross section in the non-relativistic 
approximation reads 
\begin{eqnarray}
\sigma_{eN} =
\sigma_{\rm Mott} \, 
\left[  v_L (G_E)^2 + v_T \frac{Q^2 (G_M)^2}{2 m^2} \right] 
{1 \over { 1 + \frac{2 E_e}{m} \sin^2 \frac{\theta_e}{2} } } \ .
\label{eq12}
\end{eqnarray}
(Note we always keep the kinematical factors related to the electron
relativistically).
The central question we want to answer in this paper is, how
reliable that approximation is. Clearly, this will depend on
the kinematic regime. Here we shall restrict ourselves
to photon energies $\omega$ and momenta $Q = \mid {\vec Q} \mid$ 
such that the 3N c.m. energy
in the final state is essentially below the pion mass $m_\pi$:
\begin{eqnarray}
E_{3N}^{c.m.} = \omega - \frac{ {\vec Q}^{\ 2} }{6 \, m}  + \epsilon_3 \le m_\pi
\label{eq13}
\end{eqnarray}
(to be exact: we consider cases with $E_{3N}^{c.m.} \le$ 150 MeV)
and $Q \le $ 600 MeV/c. That $Q$ value is in fact 
already somewhat too high to use strictly non-relativistic
kinematics and to neglect relativistic corrections in the
current and the dynamics. But we consider this small excursion
to be justified to acquire a first insight into a decline of FSI
with increasing $Q$-values. Qualitatively, we do not expect a change
of our results if relativistic structures will be incorporated.
We shall, however, not enter into the kinematic regime with even 
higher $Q$-values and / or $E_{3N}^{c.m.}$ significantly greater than $m_\pi$.

\begin{figure}[htb]
\begin{center}
\epsfig{file=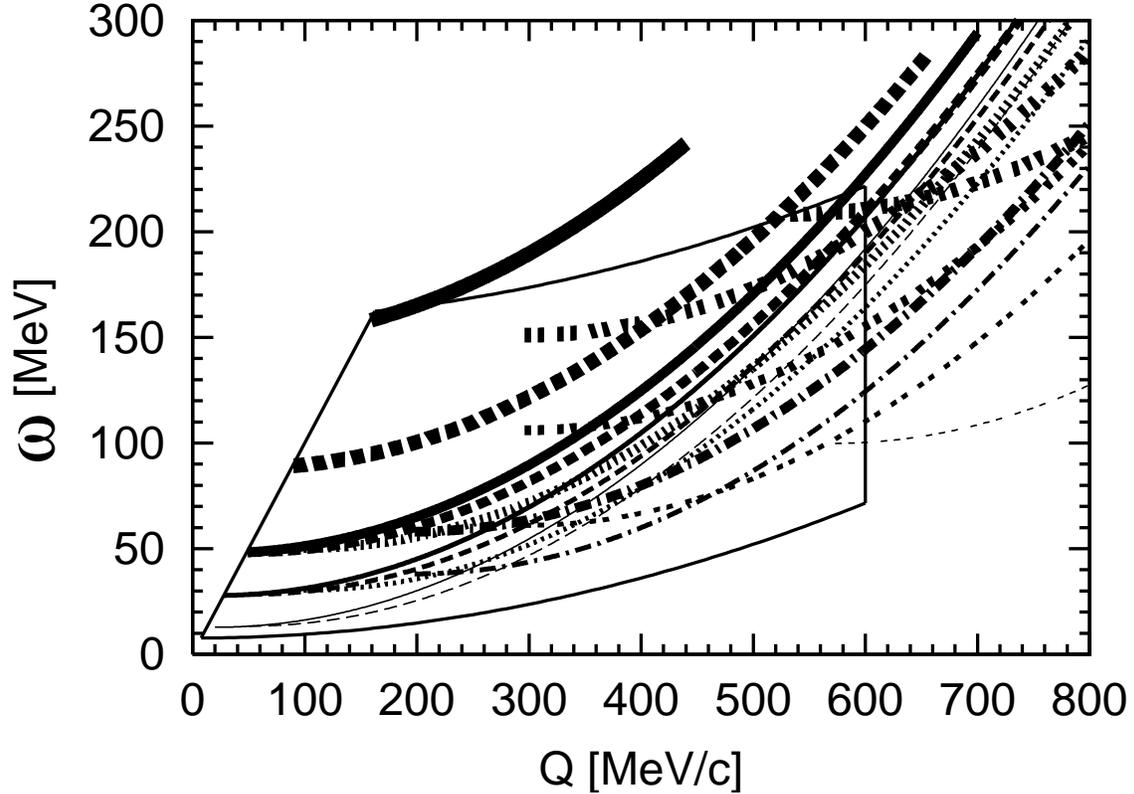,height=11cm}
\caption{\label{fig2}
The domain $D$ in the $Q-\omega$ plane for 
$E_{3N}^{c.m.} \le $ 150 MeV and $Q \le $ 600 MeV/c.
The  additional lines correspond to fixed $(k,E)$ values.
Solid lines are for $k$= 0.1 fm$^{-1}$,
dashed for $k$= 0.25 fm$^{-1}$,
dotted for $k$= 0.5 fm$^{-1}$,
dash-dotted  for $k$= 1 fm$^{-1}$,
double-dashed for $k$= 1.5 fm$^{-1}$,
and triple-dashed for $k$= 2.7-2.9 fm$^{-1}$.
The thickness of the lines increases with increasing $E$; 
it is minimal for $E$= 5 MeV  and maximal for  $E$= 140 MeV.
Note that we restrict ourselves to the ``less relativistic''
case in Eq.~(15), for which $\mid {\vec p}_1 \mid  \le \mid {\vec Q} \mid $.
%$k$= 0.1 fm$^{-1}$, $E$= 5 MeV (solid),
%$k$= 0.1 fm$^{-1}$, $E$= 140 MeV (dashed),
%$k$= 2.7 fm$^{-1}$, $E$= 125 MeV (dotted),
%$k$= 2.9 fm$^{-1}$, $E$= 5 MeV (dash-dotted),
%$k$= 1.5 fm$^{-1}$, $E$= 75 MeV (double dashed).
}
\end{center}
\end{figure}

The kinematical restriction imposed above leads to the domain $D$
in the $Q-\omega$ plane shown in Fig.~\ref{fig2}.

\begin{figure}[htb]
\begin{center}
\epsfig{file=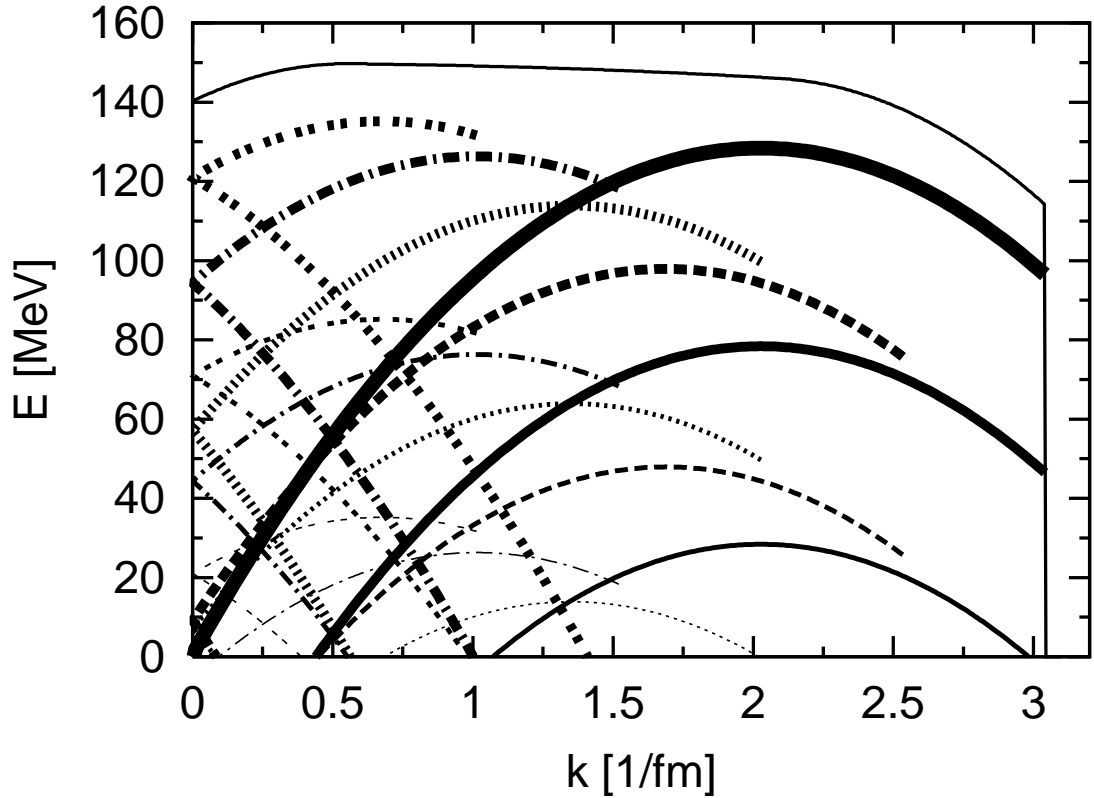,height=11cm}
\caption{\label{fig3}
The domain $D^\prime$ in the $k-E$ plane for 
$E_{3N}^{c.m.} \le $ 150 MeV and $Q \le $ 600 MeV/c.
The solid lines correspond to $Q$= 600 MeV/c,
dashed lines   to            $Q$= 500 MeV/c,
dotted lines   to           $Q$= 400 MeV/c,
dash-dotted lines to         $Q$= 300 MeV/c,
double-dotted lines to       $Q$= 200 MeV/c.
The lines thickness increases with $\omega$:
the thinest line stands for $\omega$= 50 MeV,
then the thicker and thicker lines for $\omega$= 100,
150 and 200 MeV, respectively.
%The lines correspond to fixed $(\omega,Q)$ values:
%$\omega$= 50 MeV, $Q$= 200 MeV/c (solid),
%$\omega$=150 MeV, $Q$= 200 MeV/c (dashed),
%$\omega$=100 MeV, $Q$= 600 MeV/c (dotted),
%$\omega$=200 MeV, $Q$= 600 MeV/c (dash-dotted).
      }
\end{center}
\end{figure}

Using the energy and momentum conservation in non-relativistic
kinematics leads to the following connection between the variables
$\omega$, $Q$ and  $k$, $E$:
\begin{eqnarray}
\omega + \epsilon_3 \, = \,
 \frac{(Q \pm k)^2}{2 \, m} \, + \, 
\frac{k^2}{4 \, m}  \, + \,  E \, 
\label{eq14}
\end{eqnarray}
where in (\ref{eq13}) and (\ref{eq14}) $ \epsilon_3$ is the negative
$^3$He binding energy.
The sign -(+) refers to $ 0 \le p_1 \le Q$ 
($p_1 \ge Q$), respectively. Thus taking a pair $Q-\omega$ in
$D$ provides a relation between $E$ and $k$. It is a simple matter
to map the domain $D$ into a domain $D^\prime$ in the $k-E$ plane.
This is shown in Fig.~\ref{fig3} encircled by the roughly horizontal line around 
$E$= 140 MeV and the vertical line at $k$= 3 ${\rm fm}^{-1}$.

\begin{figure}[htb]
\begin{center}
\epsfig{file=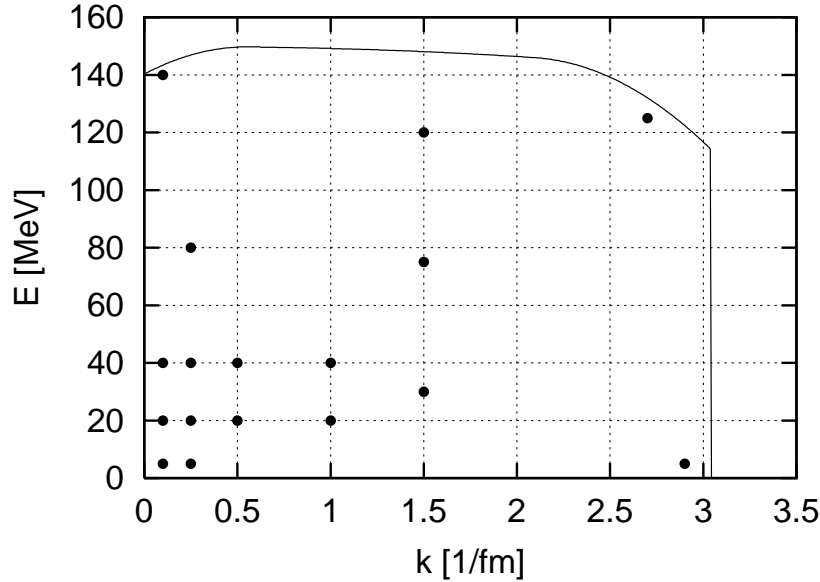,height=8cm}
\caption{\label{fig4}
The same domain $D^\prime$ shown in Fig.~\ref{fig3} in the $k-E$ plane resulting from 
$E_{3N}^{c.m.} \le $ 150 MeV and $Q \le $ 600 MeV/c
together with $(k,E)$ points for which the related $Q-\omega$ 
curves are displayed in Fig.~2.
At most of those points the validity of the FSI23 approximation
will be studied below.
For the remaining ones the validity is not given like for others shown.
      }
\end{center}
\end{figure}

To illustrate  the mappings  we also display in Fig.~3 a few examples 
for the  continuously distributed  $k-E$ pairs
to each fixed $Q-\omega$ out of $D$. We see that for fixed $Q$ 
the sequence of curves shifts upwards and to the left with 
increasing $\omega$. 
Once the bended curves hit the $k=0$ axis there
appears a branch related to the sign of $k$ in 
Eq.~(15) reaching again to nonzero $k$-values. As will be clear
below we are especially interested in the $Q-\omega$ pairs which lead to curves 
in the $k-E$ plane ending up near $E \approx 0 $ and $ k \approx 0$.

As is obvious from Eq.~(\ref{eq14}) that mapping
from $D$ to $D^\prime$ is not one-to-one. 

Thus, 
for each $k-E$ pair, only a relation between 
$Q$ and $\omega$ is determined.
Again quite a few examples are displayed in
Fig.~\ref{fig2}. Those $Q-\omega$ reach outside the domain $D$, 
where a relativistic treatment is obligatory and therefore outside 
the scope of this paper. For a better
orientation of the reader, 
%which will turn out to be useful below, 
we show the chosen $k-E$ pairs in Fig.~4.

In order to investigate the usefulness of $S$ one can use Eq.(\ref{eq9}) and
replace the response functions $R_L$, $R_T$ evaluated under the
simplifying assumption FSI23 by the full response functions
taking FSI completely into account. This is required for the 
cross section given in Eq.~(1).
Let us call the resulting expressions
$S_L (Full)$ and $S_T (Full)$, respectively. 
It is also of interest to neglect any FSI 
but keep all three terms in row 1 of Fig.~\ref{fig1}. This we call 
the symmetrized plane wave
impulse approximation, PWIAS, since then antisymmetrization is fully
taken into account, and the resulting quantities will be denoted as
$S_L (PWIAS)$ and $S_T (PWIAS)$. Finally,  
one can assume only the very first process in Fig.~\ref{fig1}
to be present, leading to $S_L (PWIA)$ and $S_T (PWIA)$. 
In this  manner we can compare
$S(k,E)$ to the other three choices of dynamical input. Each $k-E$ pair
fixes according to Eq.~(\ref{eq14}) $\omega$ if $Q$ is given. Thus we shall plot
the four $S$'s for fixed $(k,E)$ as a function of $Q$;
in other words as a function of the electron kinematics.
By construction $S_L (PWIA)$, $S_T (PWIA)$ and $S$ are functions of $k$ and $E$ 
only and do not depend on $Q$.
This, however, does not hold for 
$S_L (PWIAS)$ and $S_T (PWIAS)$ and the results
based on full treatment of FSI, $S_L (Full)$ and $S_T (Full)$.

Obviously Eq.~(\ref{eq14}) can also be written as
\begin{eqnarray}
\omega + \epsilon_3 \, = \,
 E_1 \, + \, \frac{k^2}{4 \, m}  \, + \,  E \, 
\label{eq15}
\end{eqnarray}
with $E_1 = \frac{p_1^2}{2m}$. Therefore $E_1$ 
can be equally used as the abscissa.
Note, however, that the different ${E_1}'s$  belong 
to different electron kinematics. This is one way to represent our results
starting from fixed $(k,E)$ values. We shall also provide examples using
a fixed electron kinematics and plot the results as a function of $E_1$, 
which is more natural in relation to the experiment.

\section{\label{sec3}Results}

In all calculations the AV18 nucleon-nucleon potential \cite{AV18} has been used
without its electromagnetic parts.
It is plausible to assume for the parallel kinematics
considered in the paper that meson exchange currents (MEC) do not play 
any essential role. Thus we concentrate on the FSI effects and neglect
any contribution from MEC.

Under the simplifying assumptions represented by the two encircled diagrams
in Fig.~\ref{fig1}, the response functions $R_L$ and $R_T$ 
%which determine  the cross section in parallel kinematics, 
are directly linked to the spectral function $S$, as shown in Eq.~(10).
In order to achieve insight under
which conditions this form has validity, we shall cover the domain $D^\prime$
in Fig.~\ref{fig3} by a representative grid of $(k,E)$ points chosen in Fig.~\ref{fig2} and
marked by dots in Fig.~\ref{fig4}. 
%????? WE DO NEED MORE !       YES

\begin{figure}[htb]
\begin{center}
\epsfig{file=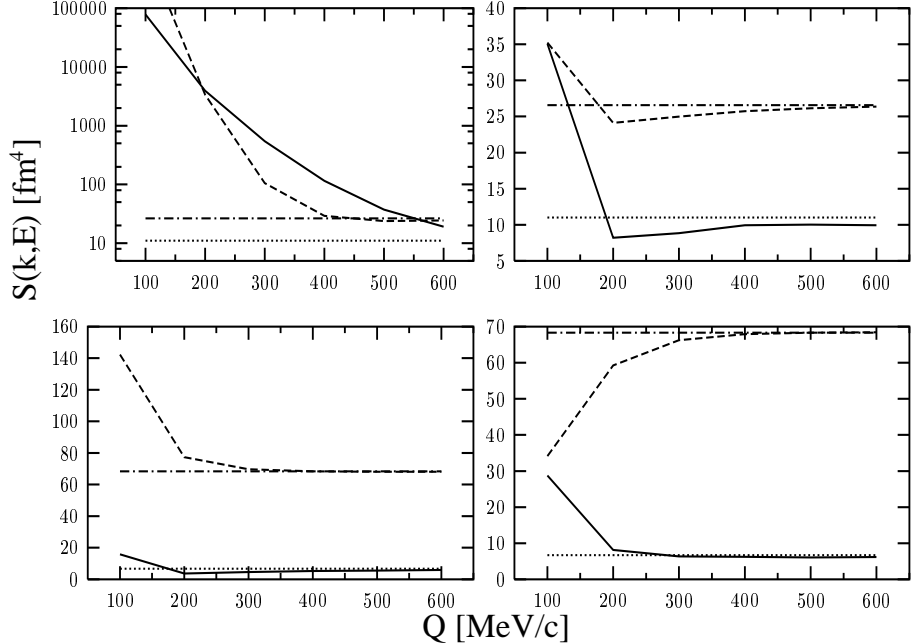,height=9cm}
\caption{\label{fig5}
The spectral function $S(k,E)$ 
and results based on the form given in  Eq.~(\ref{eq9}) 
but  using different dynamical assumptions 
for the response functions $R_L$ and $R_T$ as a function of the momentum transfer $Q$ 
for a fixed 
$(k,E)$ pair:  $k$= 0.1 fm$^{-1}$, $E$= 5 MeV. 
Top figures describe the neutron knockout and bottom ones 
the proton case. The longitudinal (left figures) 
and transverse (right figures) response functions  
are employed.
PWIA (dash-dotted line), PWIAS (dashed) 
and Full results (solid line) are shown.
The FSI23 result (dotted)  is the spectral function $S(k,E)$,
which is independent of $Q$.  
}
\end{center}
\end{figure}

\begin{figure}[htb]
\begin{center}
\epsfig{file=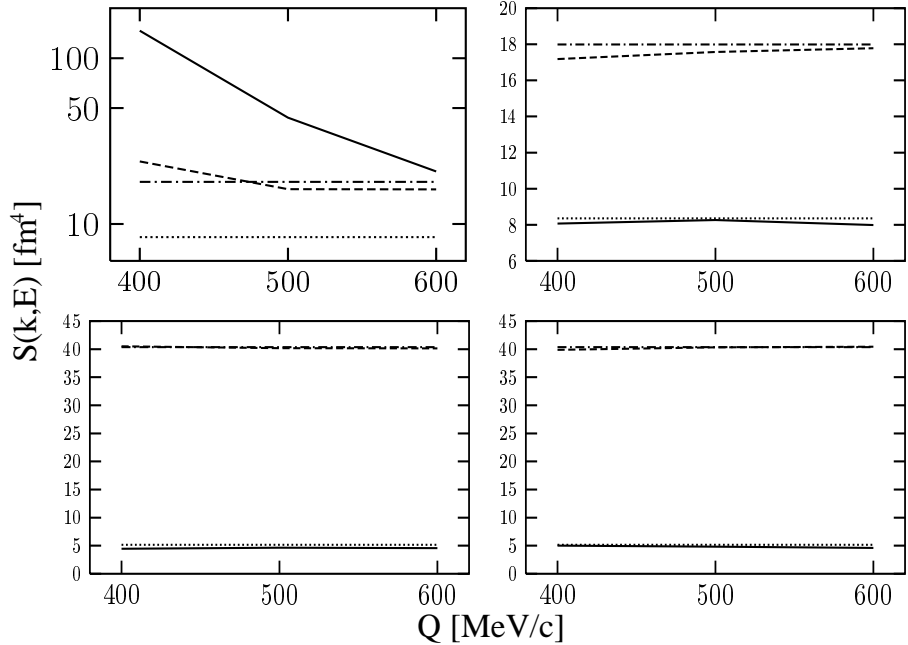,height=9cm}
\caption{\label{fig6}
The same as in Fig.~\ref{fig5}
for $k$= 0.25 fm$^{-1}$, $E$= 5 MeV.
      }
\end{center}
\end{figure}

\begin{figure}[htb]
\begin{center}
\epsfig{file=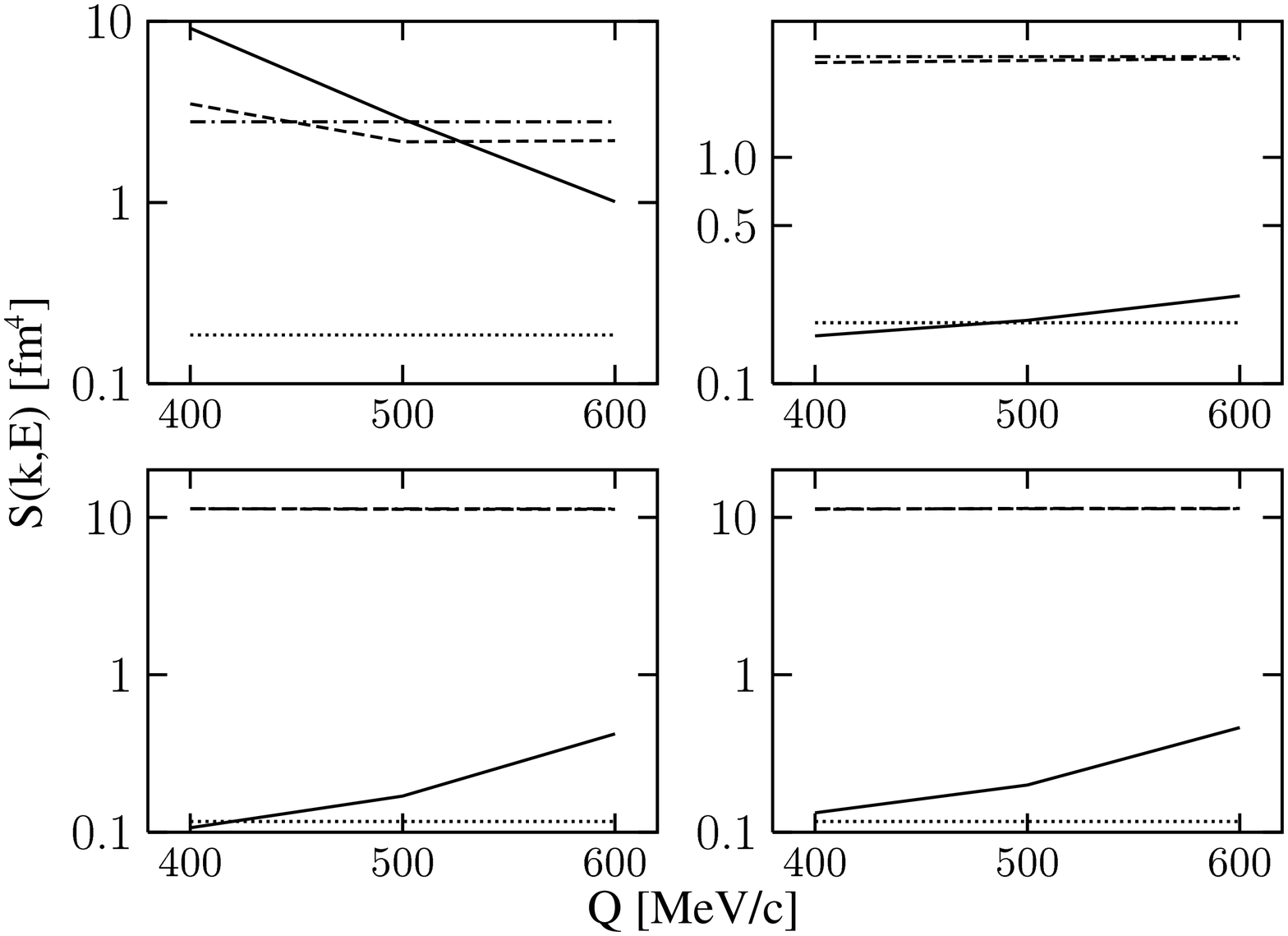,height=9cm}
\caption{\label{fig7}
The same as in Fig.~\ref{fig5}
for $k$= 0.1 fm$^{-1}$, $E$= 20 MeV.
      }
\end{center}
\end{figure}

\begin{figure}[htb]
\begin{center}
\epsfig{file=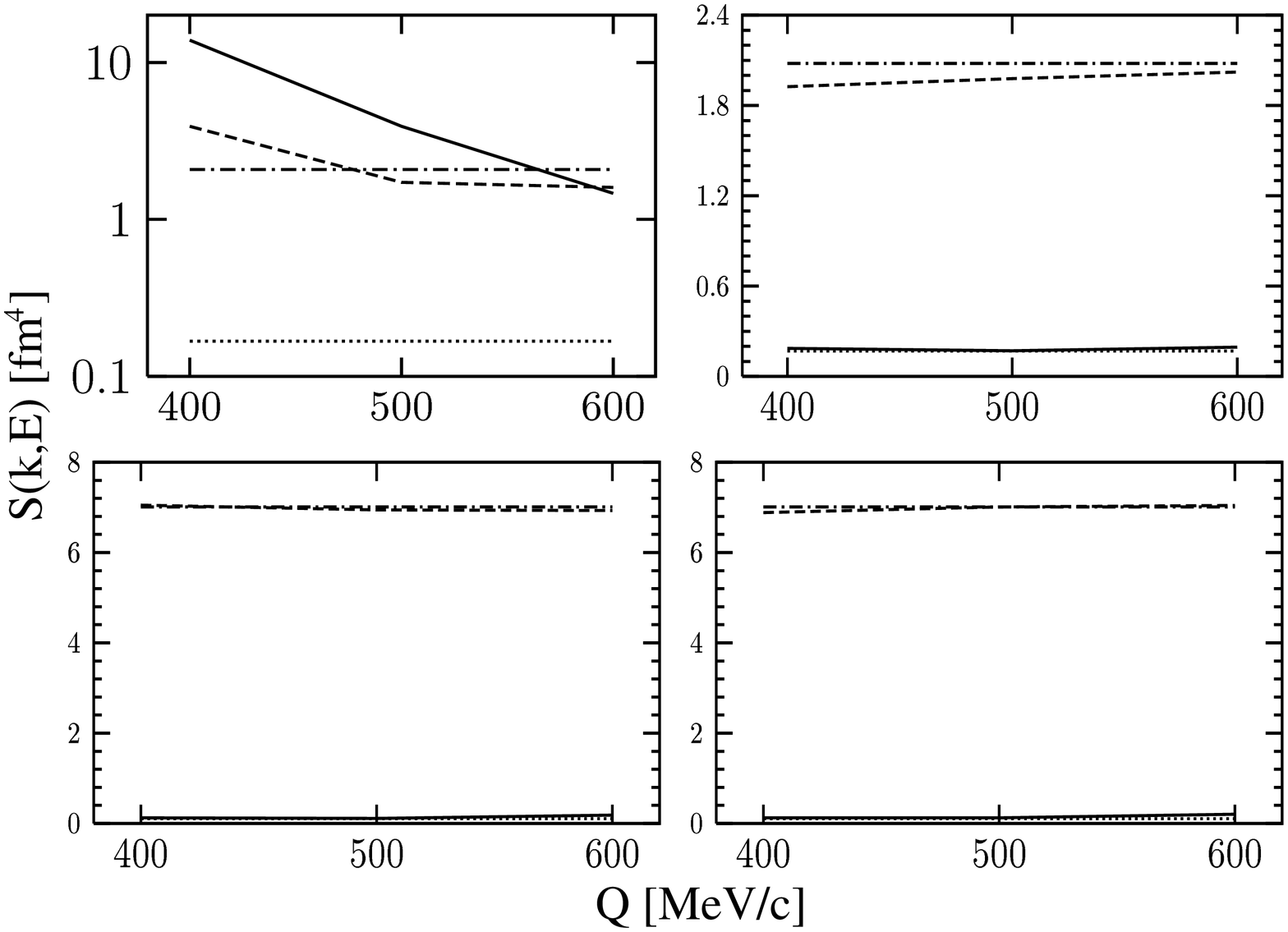,height=9cm}
\caption{\label{fig8}
The same as in Fig.~\ref{fig4}
for $k$= 0.25 fm$^{-1}$, $E$= 20 MeV.
      }
\end{center}
\end{figure}

\begin{figure}[htb]
\begin{center}
\epsfig{file=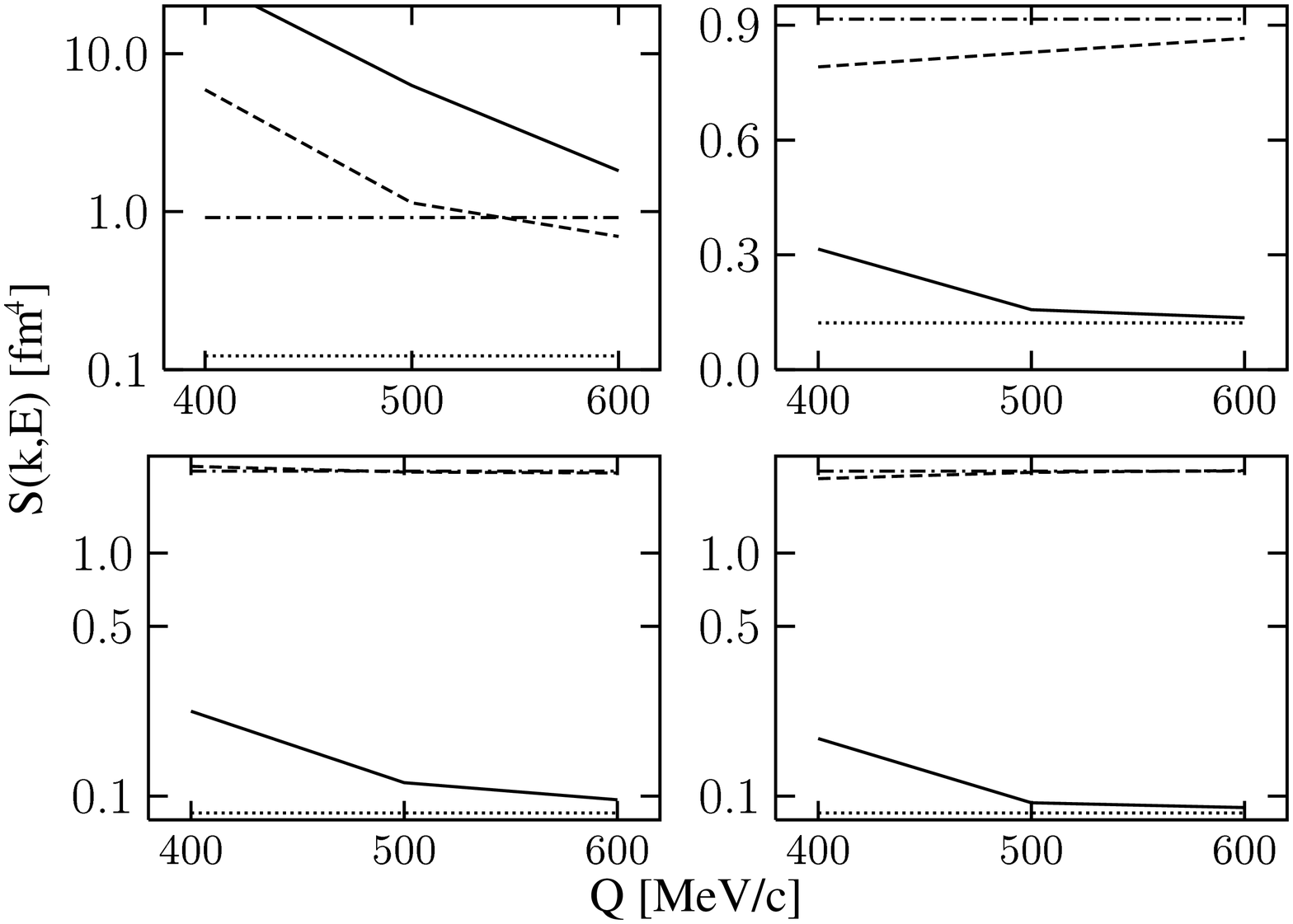,height=9cm}
\caption{\label{fig9}
The same as in Fig.~\ref{fig5}
for $k$= 0.5 fm$^{-1}$, $E$= 20 MeV.
      }
\end{center}
\end{figure}

\begin{figure}[htb]
\begin{center}
\epsfig{file=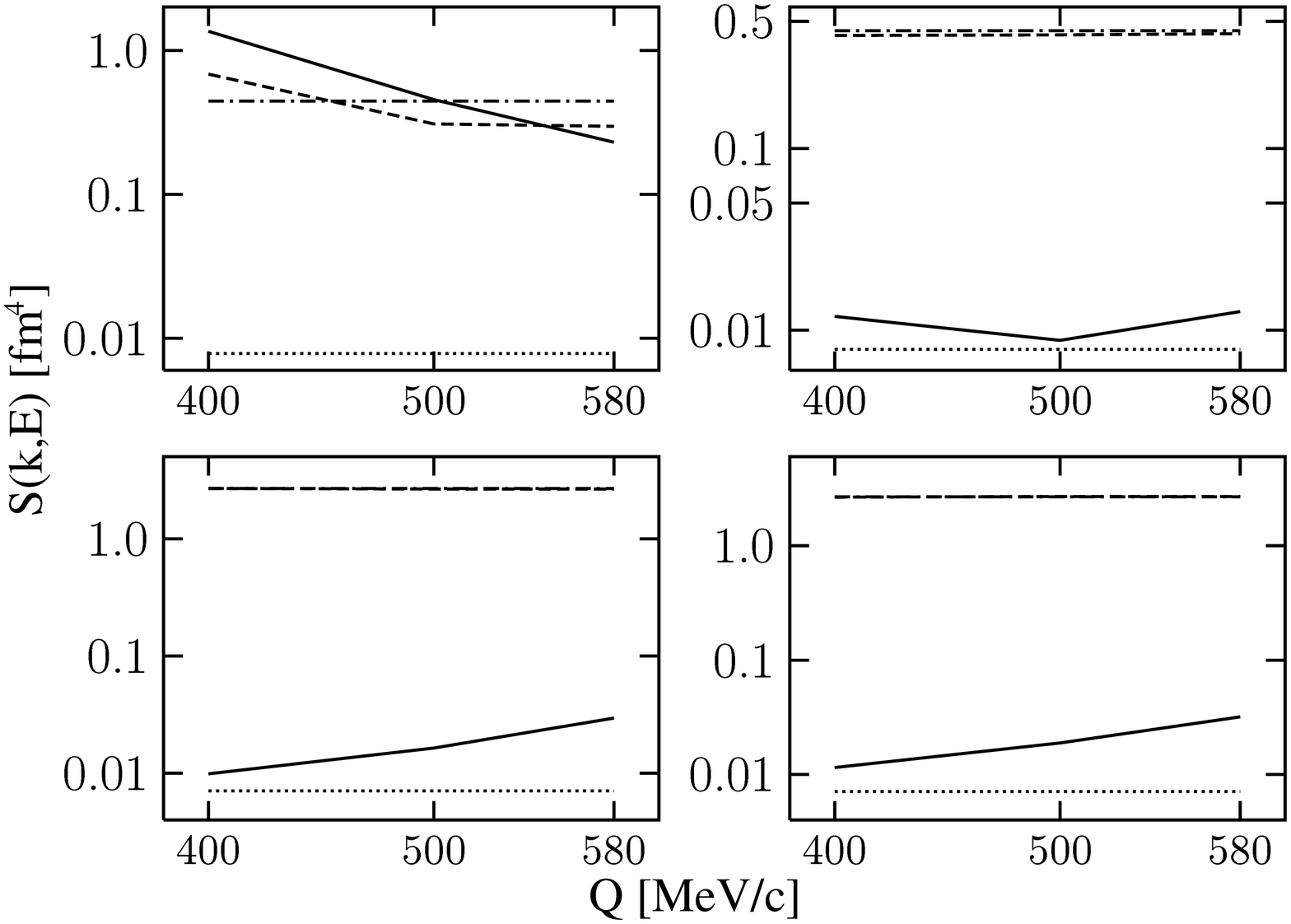,height=9cm}
\caption{\label{fig10}
The same as in Fig.~\ref{fig5}
for $k$= 0.1 fm$^{-1}$, $E$= 40 MeV.
      }
\end{center}
\end{figure}

\begin{figure}[htb]
\begin{center}
\epsfig{file=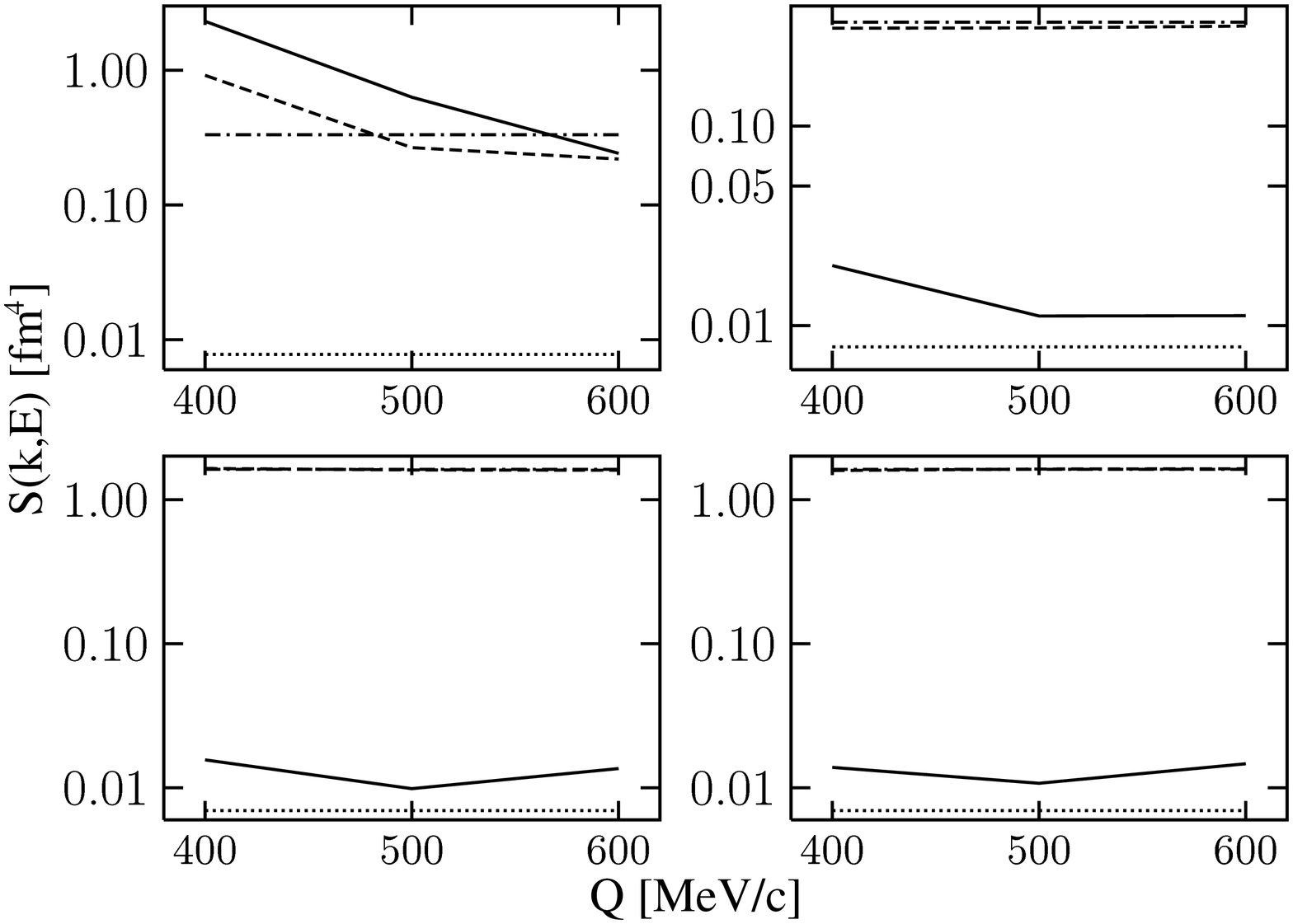,height=9cm}
\caption{\label{fig11}
The same as in Fig.~\ref{fig5}
for $k$= 0.25 fm$^{-1}$, $E$= 40 MeV.
      }
\end{center}
\end{figure}

\begin{figure}[htb]
\begin{center}
\epsfig{file=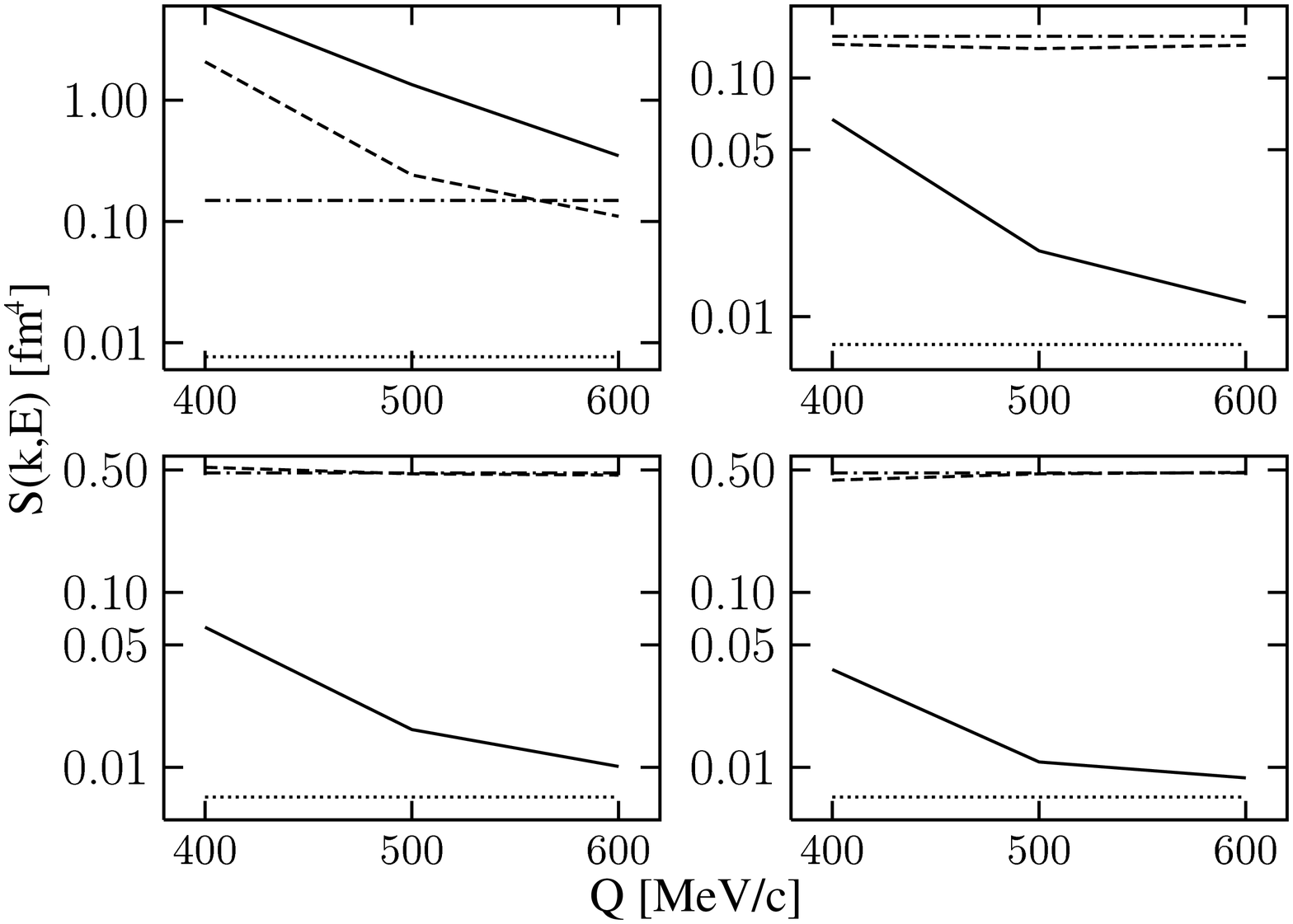,height=9cm}
\caption{\label{fig12}
The same as in Fig.~\ref{fig5}
for $k$= 0.5 fm$^{-1}$, $E$= 40 MeV.
      }
\end{center}
\end{figure}

\begin{figure}[htb]
\begin{center}
\epsfig{file=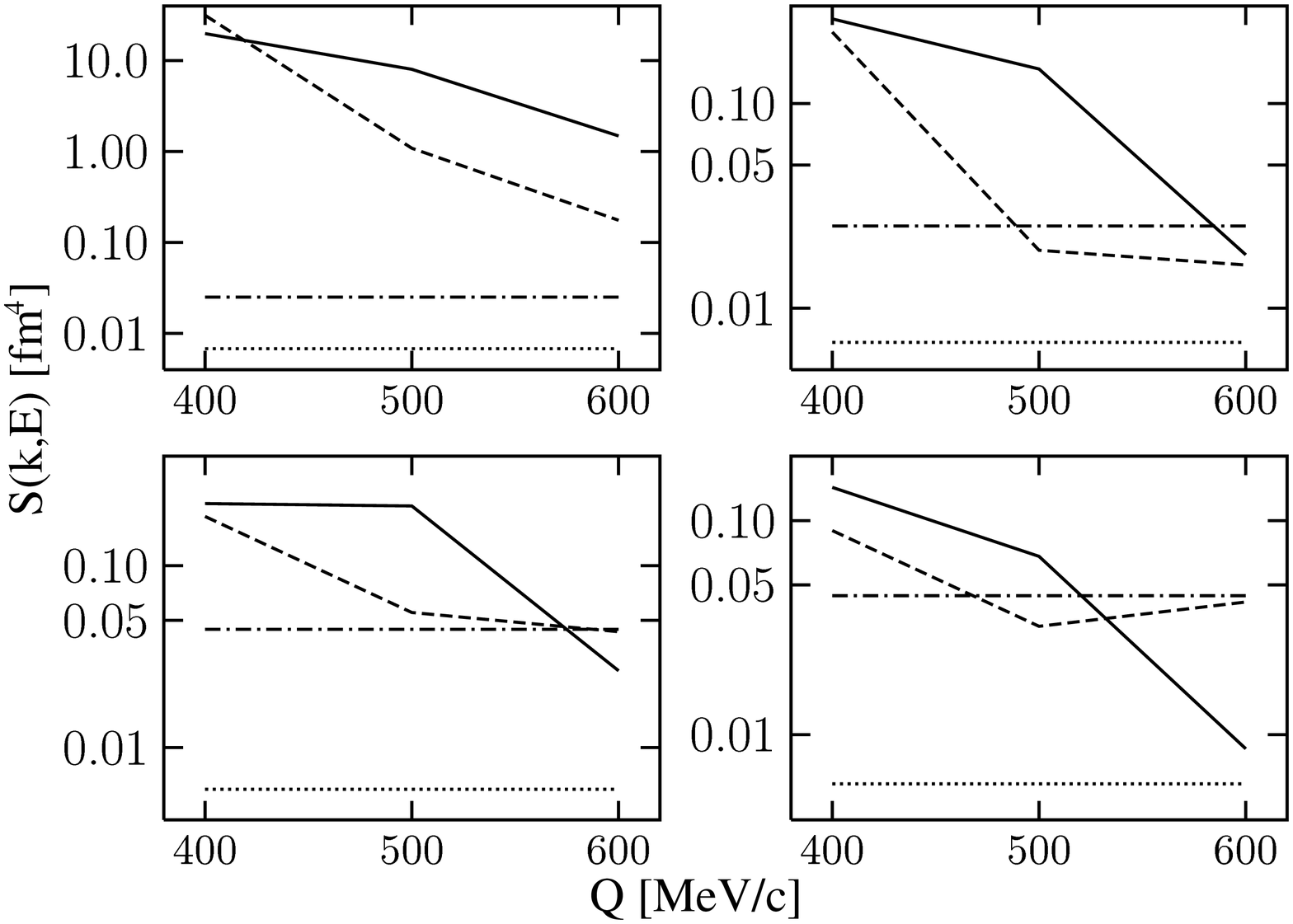,height=9cm}
\caption{\label{fig13}
The same as in Fig.~\ref{fig5}
for $k$= 1.0 fm$^{-1}$, $E$= 40 MeV.
      }
\end{center}
\end{figure}

\begin{figure}[htb]
\begin{center}
\epsfig{file=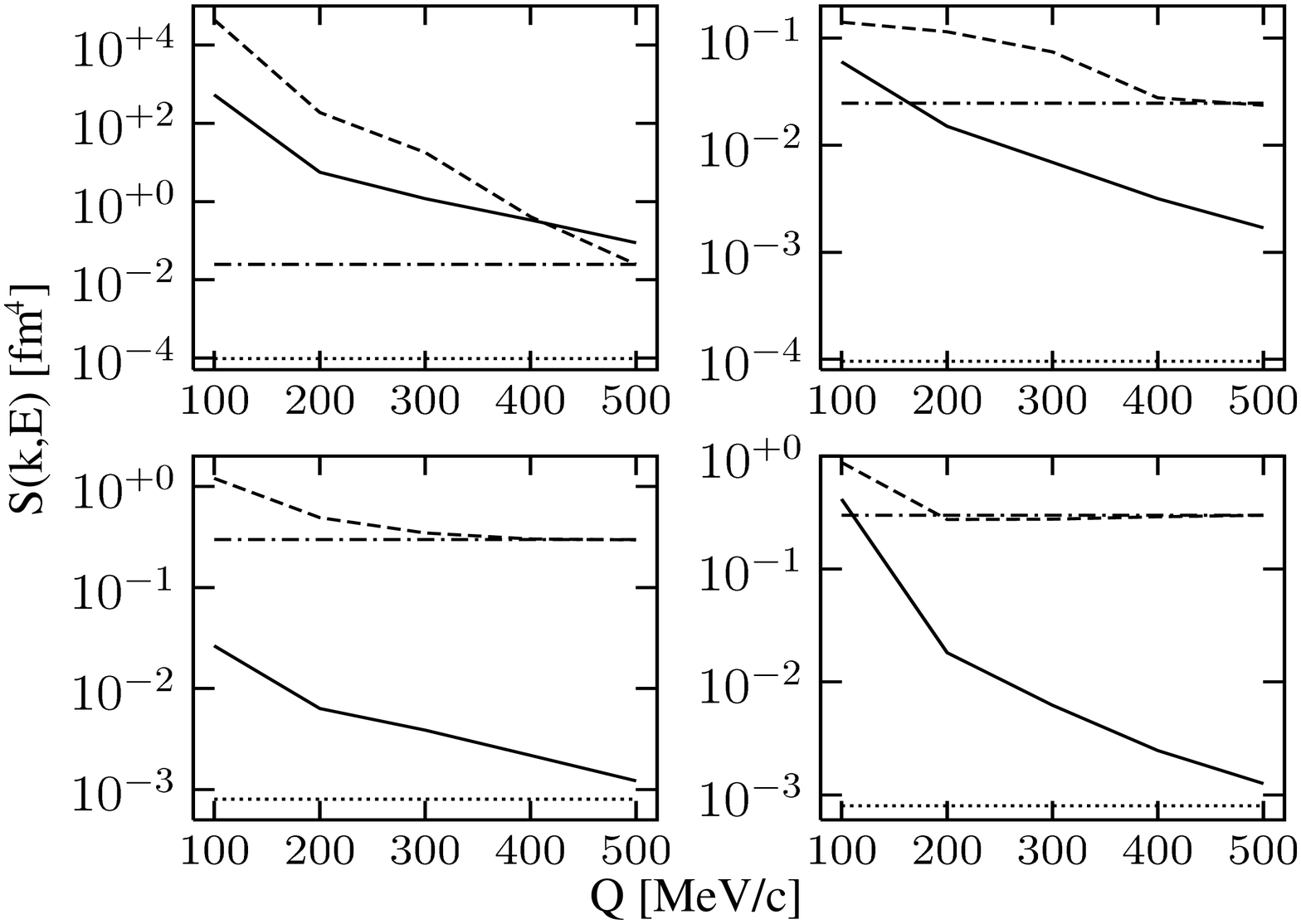,height=9cm}
\caption{\label{fig14}
The same as in Fig.~\ref{fig5}
for $k$= 0.25 fm$^{-1}$, $E$= 80 MeV.
      }
\end{center}
\end{figure}

\begin{figure}[htb]
\begin{center}
\epsfig{file=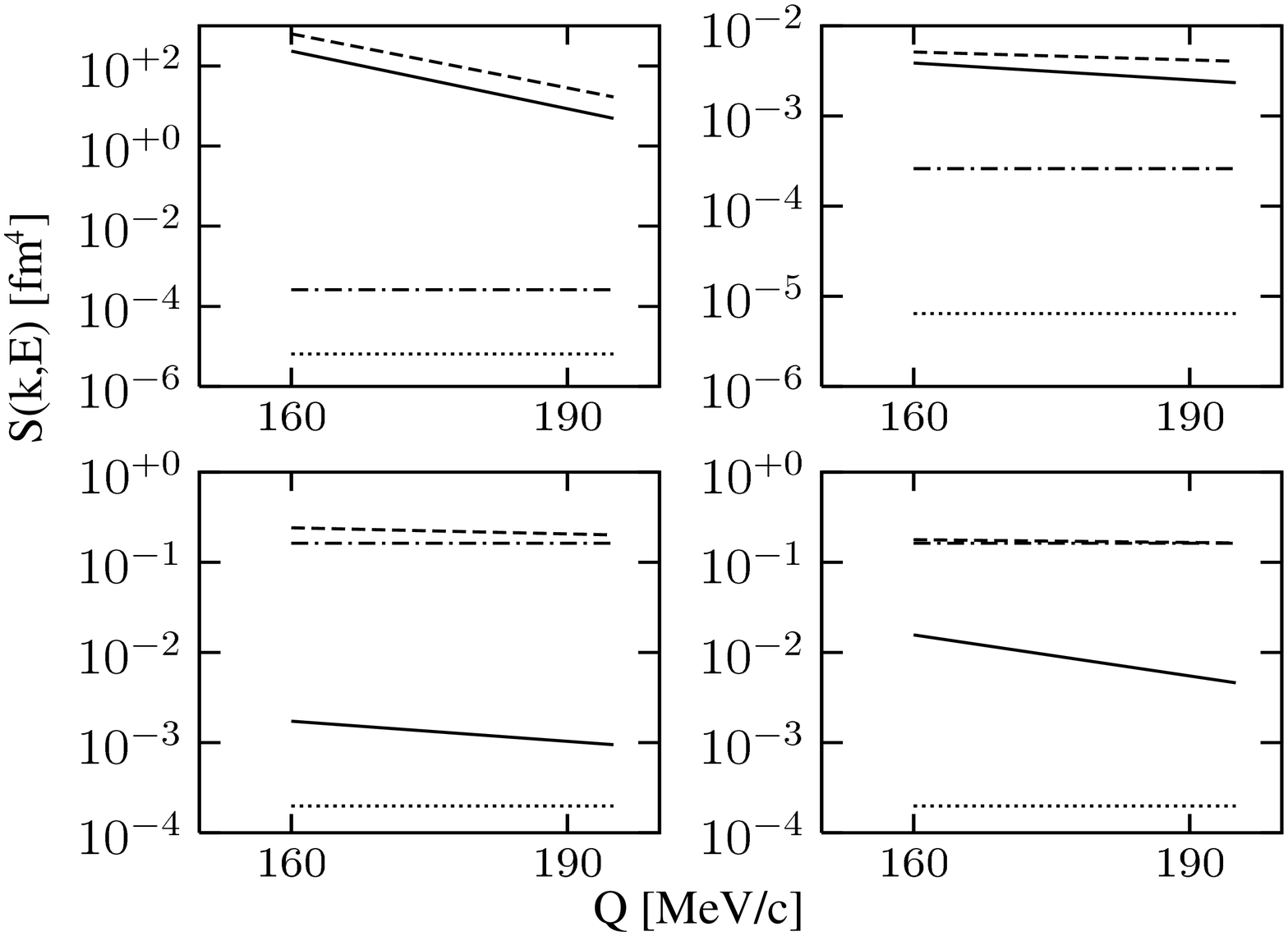,height=9cm}
\caption{\label{fig15}
The same as in Fig.~\ref{fig5} 
for $k$= 0.1 fm$^{-1}$, $E$= 140 MeV.
      }
\end{center}
\end{figure}

\begin{figure}[htb]
\begin{center}
\epsfig{file=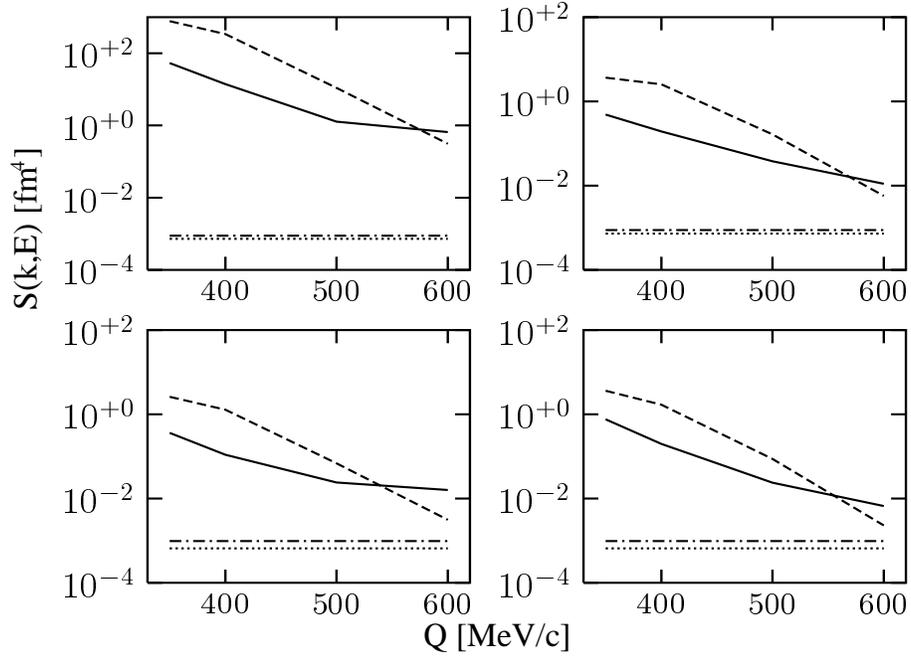,height=9cm}
\caption{\label{fig16}
The same as in Fig.~\ref{fig5} 
for $k$= 1.5 fm$^{-1}$, $E$= 75 MeV.
      }
\end{center}
\end{figure}

\begin{figure}[htb]
\begin{center}
\epsfig{file=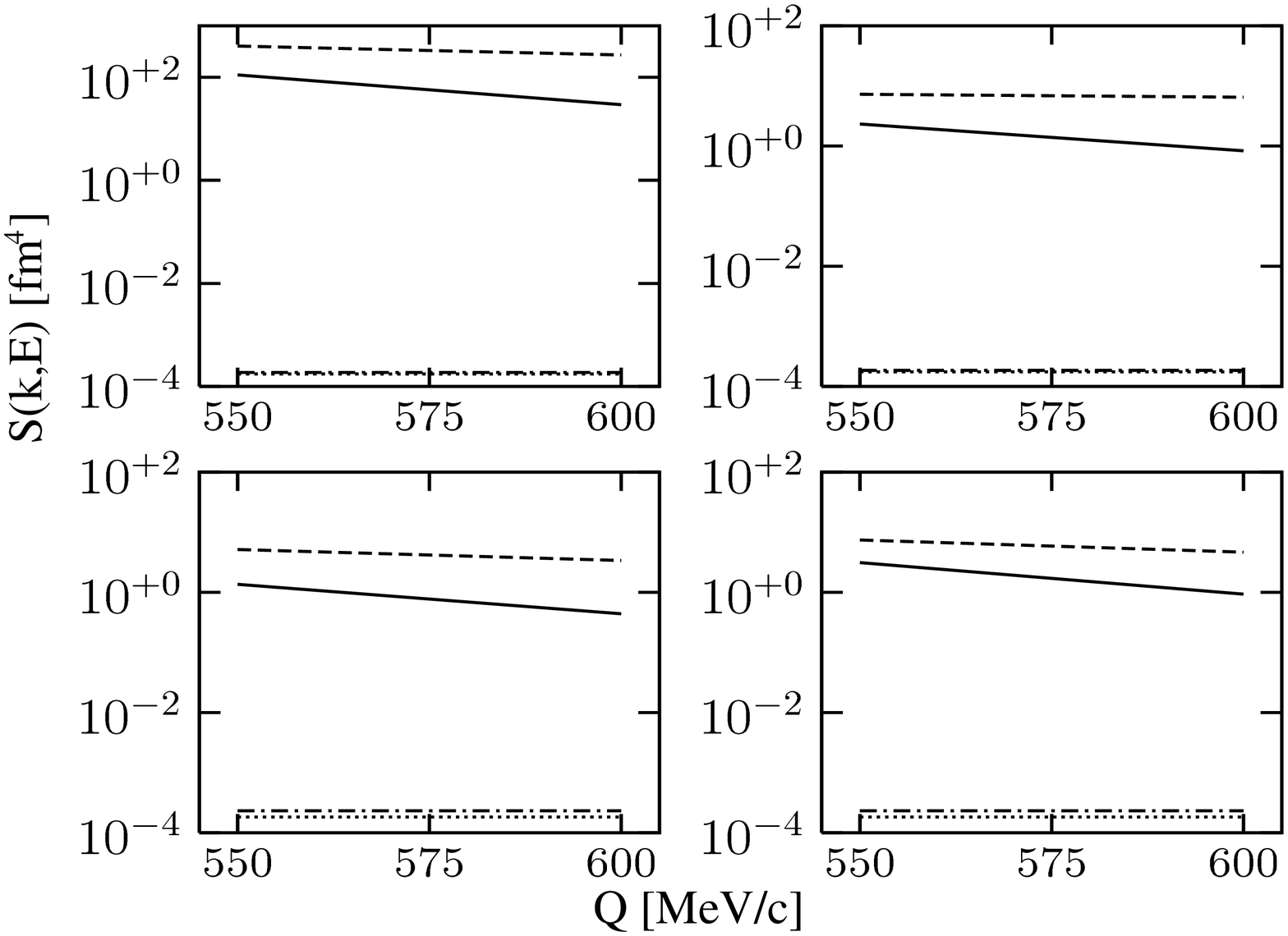,height=9cm}
\caption{\label{fig17}
The same as in Fig.~\ref{fig5} 
for $k$= 2.7 fm$^{-1}$, $E$= 125 MeV.
      }
\end{center}
\end{figure}

To each such
pair corresponds a quadratic relation between the photon energy
$\omega$ and its three-momentum $Q$, as given in Eq.~(\ref{eq14}). 
This traces out
a curve and examples thereof are shown in Fig.~\ref{fig2}. We shall now choose
those curves inside the domain $D$ and compare
the spectral function $S$ to the expressions $S_L (Full)$ and $S_T (Full)$
evaluated under the full sequence of rescattering processes, and further compare 
$S$ to $S_L(PWIAS)$ and $S_T(PWIAS)$
taking the correct antisymmetrization into account but neglecting any final state
interaction and finally  we compare $S$ to $S_L(PWIA)$ and $S_T(PWIA)$
keeping only the very first process in
Fig.~\ref{fig1}. The results are displayed
in Figs.~\ref{fig5}--\ref{fig17}.

%\begin{figure}[htb]
%\begin{center}
%\epsfig{file=IV.eps,height=9cm}
%\caption{\label{fig7}
%The same as in Fig.~\ref{fig4} 
%for $k$= 2.9 fm$^{-1}$, $E$= 5 MeV.
%      }
%\end{center}
%\end{figure}
%
%\begin{figure}[htb]
%\begin{center}
%\epsfig{file=VII.eps,height=9cm}
%\caption{\label{fig10}
%The same as in Fig.~\ref{fig4}
%for $k$= 1.5 fm$^{-1}$, $E$= 30 MeV.
%      }
%\end{center}
%\end{figure}
%
%
%\begin{figure}[htb]
%\begin{center}
%\epsfig{file=VIII.eps,height=9cm}
%\caption{\label{fig11}
%The same as in Fig.~\ref{fig4}
%for $k$= 1.5 fm$^{-1}$, $E$= 120 MeV.
%      }
%\end{center}
%\end{figure}
%
%
%\begin{figure}[htb]
%\begin{center}
%\epsfig{file=XI.eps,height=9cm}
%\caption{\label{fig14}
%The same as in Fig.~\ref{fig4}
%for $k$= 1.0 fm$^{-1}$, $E$= 20 MeV.
%      }
%\end{center}
%\end{figure}
%

%
%\begin{figure}[htb]
%\begin{center}
%\epsfig{file=XII.eps,height=9cm}
%\caption{\label{fig15}
%The same as in Fig.~\ref{fig4}
%for $k$= 0.5 fm$^{-1}$, $E$= 40 MeV.
%      }
%\end{center}
%\end{figure}
%

% .... for $R_L$ and in Figs ... for $R_T$. The abzissa shows
%$Q$ and the corresponding $\omega$ values. 
%According to Fig. () also the energy
%of the knocked out proton $E_1$ is uniquely given and shown as a third
%information on the abzissa. 
Let us first concentrate on the full calculation represented by a solid line 
in comparison to  the spectral function $S$ given 
as a dotted line in case of the
proton knockout process (lower panels).  We see that
only for  the $k-E$ pairs (0.1 fm$^{-1}$, 5 MeV), 
(0.25 fm$^{-1}$, 5 MeV), (0.25 fm$^{-1}$, 20 MeV), 
(0.5 fm$^{-1}$, 20 MeV) and (0.5 fm$^{-1}$, 40 MeV) chosen in Fig.~4,
the two curves approach each other with increasing 
$Q$-values within the considered range of $Q$-values.

Though the cases 
($k$=0.1  fm$^{-1}$, $E$=20 MeV), 
($k$=0.1  fm$^{-1}$, $E$=40 MeV) and 
($k$=0.25 fm$^{-1}$, $E$=40 MeV) also are small nearby pairs,
the two curves do not approach each other. We have no explanation 
for that unexpected behavior.
This means that one should stay on the safe side and better check even
if both $k$ and $E$ approach small values whether the FSI of the knocked
out nucleon is really negligible.

%Jetzt muss ein Kommentar folgen, warum  
%fuer (0.1,20),(0.1,40) (0.25,40) dies nicht
%funktioniert. Auch  wenn ich die Kinematik des Files I-XVII anschaue, 
%kommt mir keine Idee.  Verstehen Sie dies?  

That approach is
qualitatively similar for $R_L$ and $R_T$. For the other $k-E$ pairs the FSI23 approximation
leading to the spectral function is by far not sufficient and the full rescattering takes place. We also show the very first
process
in Fig.~\ref{fig1} denoted by PWIA and a second case  where the correct
antisymmetrization is kept but no rescattering process is allowed.
This we denote by PWIAS. Figures~\ref{fig5}-\ref{fig17} 
exhibit different situations in relation 
of the PWIAS versus
the PWIA results and the PWIA versus the FSI23 results. 
In nearly all cases shown
symmetrization in plane wave approximation (PWIAS) is quite unimportant except sometimes
at the small $Q$-values. In the cases (1.5 fm$^{-1}$, 75 MeV) 
and (2.7 fm$^{-1}$, 125 MeV) symmetrization, 
however, is quite important.
All that is easily understood regarding the momentum values for the two additional processes of
PWIAS. In the case of PWIA the $^3$He wave function $ \Psi_{{}^3{\rm He}} ( {\vec p} ,  {\vec q} ) $
is evaluated for $ {\vec p} = \frac12 \left( {\vec p}_2 - {\vec p}_3 \right) $
and $ {\vec q} = {\vec p}_1 - {\vec Q}$. 
For the two additional processes present in PWIAS 
the corresponding arguments are 
$\left(  {\vec p} = \frac12 \left( {\vec p}_1 - {\vec p}_2 \right) ,
  {\vec q} = {\vec p}_3 - {\vec Q} \right)$
and
$\left(  {\vec p} = \frac12 \left( {\vec p}_3 - {\vec p}_1 \right) ,
  {\vec q} = {\vec p}_2 - {\vec Q} \right)$.
%    ( KANN MAN dies klar machen, indem man die Impulsargumente  algebraisch angibt??).
Interestingly in case of Figs.~\ref{fig16} and~\ref{fig17} 
the PWIA and FSI23 results agree very well.
Thus the final state interaction among the two spectator nucleons is negligible. 
If additionally the symmetrization 
and all of the final state interaction were negligible, one would
have a perfect view right away into the $^3$He wave function, since $S$ evaluated under PWIA
condition displays directly the magnitude of the $^3$He wave function. (This is obvious
from Eq.~(7) if one drops the contribution proportional to  $t_{23} G_0$). 
That neglection is, however, not justified as documented 
in Figs.~\ref{fig16} and~\ref{fig17}. 
Already the correct antisymmetrization, which is
independent of FSI changes the results totally. 

In \cite{ciof02} $S(k,E)$ is displayed together with $S$ evaluated under the 
PWIA condition. 
They essentially agree for $k \ge 1.5 \ {\rm fm}^{-1}$ along $E=k^2/(4m)$.
As examples one could take $k= 2 \ {\rm fm}^{-1}$ 
corresponding to $E \approx 40$ MeV or 
$k= 3 \ {\rm fm}^{-1}$ 
with $E$ about $90$ MeV. This suggested direct insight into the $^3$He wave function.
In view of our results shown in Figs.~\ref{fig16} and \ref{fig17} 
%and \ref{fig11} this
this suggestion is not valid if the electron kinematics belongs to the domain $D$. 

%%%??? I HAVE TO CHECK THE VALUES of k and E where it happens.
%%This was shown also by Kievsky, for example. 

%( fuer k,E gegen Null ist es vielleicht gueltig??)
%evaluated under PWIA condition displays directly the magnitude of
%the  wave function. Now we see from Figs.~\ref{fig9}--~\ref{fig10} 
%that for those $(k-E)$
%values the correct antisymmetrization alone excludes the
%applicability of that fact. The additional terms where the photon
%is absorbed by the other two  nucleons inside $^3$He also contribute
%strongly to the process for all $\omega$-$Q$ pairs in the domain $D$.
%On top, of course, the full FSI is not at all negligible for those
%$k-E$ pairs. Thus inside the domain $D$ that view directly into $^3$He
%is not possible. (What happens at k,E going towards zero ???)

This  does, of course, not exclude that outside of $D$ the situation
might be more favorable to such an ideal situation. In Fig.~\ref{fig2}
one can see that for those
pairs of $k-E$ values there are continuous $\omega-Q$ pairs, 
where such an ideal situation might
exist. This requires,
however, above all a relativistic treatment and taking all the
additional dynamical ingredients into account, which is outside
the scope of the present study. Please also note that for c.m. 3N energies
above the pion threshold no nuclear forces comparable in quality  to the
ones below are available.

Regarding now the neutron knockout even at low $k-E$ values the spectral 
function in case
of $R_L$ is insufficient. For $R_T$, however, the situation is quite 
similar to the proton knockout.
Thus neutron knockout for $^3$He without separation of $R_L$ and $R_T$
is not suitable for that application of the spectral function $S$.  

We have to conclude that for  most of the $Q-\omega$ values in the domain $D$ the use
of the spectral function is quantitatively not justified and identifying
experimentally extracted $S$-functions after integration over $E$ with
the $^3$He momentum distribution is not correct. It is only for a certain group
of very small $k-E$ values (both) and for proton knockout that $S_L (Full)$ and $S_T (Full)$
approach $S$ at the higher
$Q$ values in the domain $D$. 

%Kann man die omegas  und Q Werte, wo es funktioniert in D in
%etwa angeben?  Wenn Sie die Omegaa-Q Kurven fuer unsere Paare
%(k,E) in Fig 2 eintragen, kann man  vielleicht benennen, 
%wo dieser funktionierender
%Bereich liegt. 

It is at least that ``corner'' of the $k-E$ domain where the
theoretical prediction should be valid since 
only the $NN$ $t$-matrix together with 
the $^3$He wave function enter at low momenta. 
% (ist dies richtig??) Yes, this is OK ! 
Therefore precise data there would
be quite important to validate at least that expectation. 
For other regions inside $D$ the full dynamics is acting.
% This happens at the quasi free peak region
%and there even $S_{PWIA}$ is a sufficiently good approximation. It is only
%there that a direct view into $^3$He appears to be possible and precise
%data there appear to us very desirable to see whether at least
%in that "corner" of the $k-E$ plane the theoretical prediction is correct.

%Then in the neighborhood of that corner the dynamical  ingredients,
%FSI, MEC and 3NF's will start to play their roles and systematic comparisons
%of theory and precise new data would be important. For the 3N system inside 
%the domain $D$ the dynamical equations can be solved reliably and therefore disagreement
%with the data reveals unambiguous deficiencies in all or some  of the various
%dynamical ingredients.

In actual experiments it is natural to present the data for the process
${}^3{\rm He}( e,e'N)$ 
for a given $Q-\omega$ pair as a function of $E_1$, the energy of 
the knocked out nucleon. In this case the $k-E$ values trace out a curve 
in the domain $D^\prime$ as shown  in  Fig.~\ref{fig3}. Of course investigating such a scenario
there will be no new information beyond the one we already displayed.
Nevertheless since this is what appears naturally in an experiment
we would like to show the corresponding $S$-curves now as a function of $E_1$.
First we choose proton knockout and take $Q-\omega$ values which in the $k-E$
plane lead to curves ending up in the "corner", 
where both $k$ and $E$ are rather small. As
seen from Eq.~(15) and displayed in some examples 
in Fig.~\ref{fig3} suitable cases are: 
%$\omega =  50$ MeV and $Q= 200, 300$ MeV/c; 
$\omega = 100$ MeV,    $Q= 400$ MeV/c; 
$\omega = 100$ MeV,    $Q= 500$ MeV/c; 
%$\omega = 150$ MeV,    $Q= 500$ MeV/c; 
$\omega = 150$ MeV,    $Q= 600$ MeV/c.
%$\omega = 200$ MeV,    $Q= 600$ MeV/c.

%Sollen wir den alle Faelle nehmen? 
%zum Beispiel 50,200 ist nicht so toll. Liegt
%wahrscheinlich daran, dass man zu weit links 
%landet (E ist groesser Null fuer k=0) .
%Um auch etwas Platz zu sparen, schlage ich  vor nur zu nehmen:
%
%    100,400
%    100,500
%    150,600 

\begin{figure}[htb]
\begin{center}
\epsfig{file=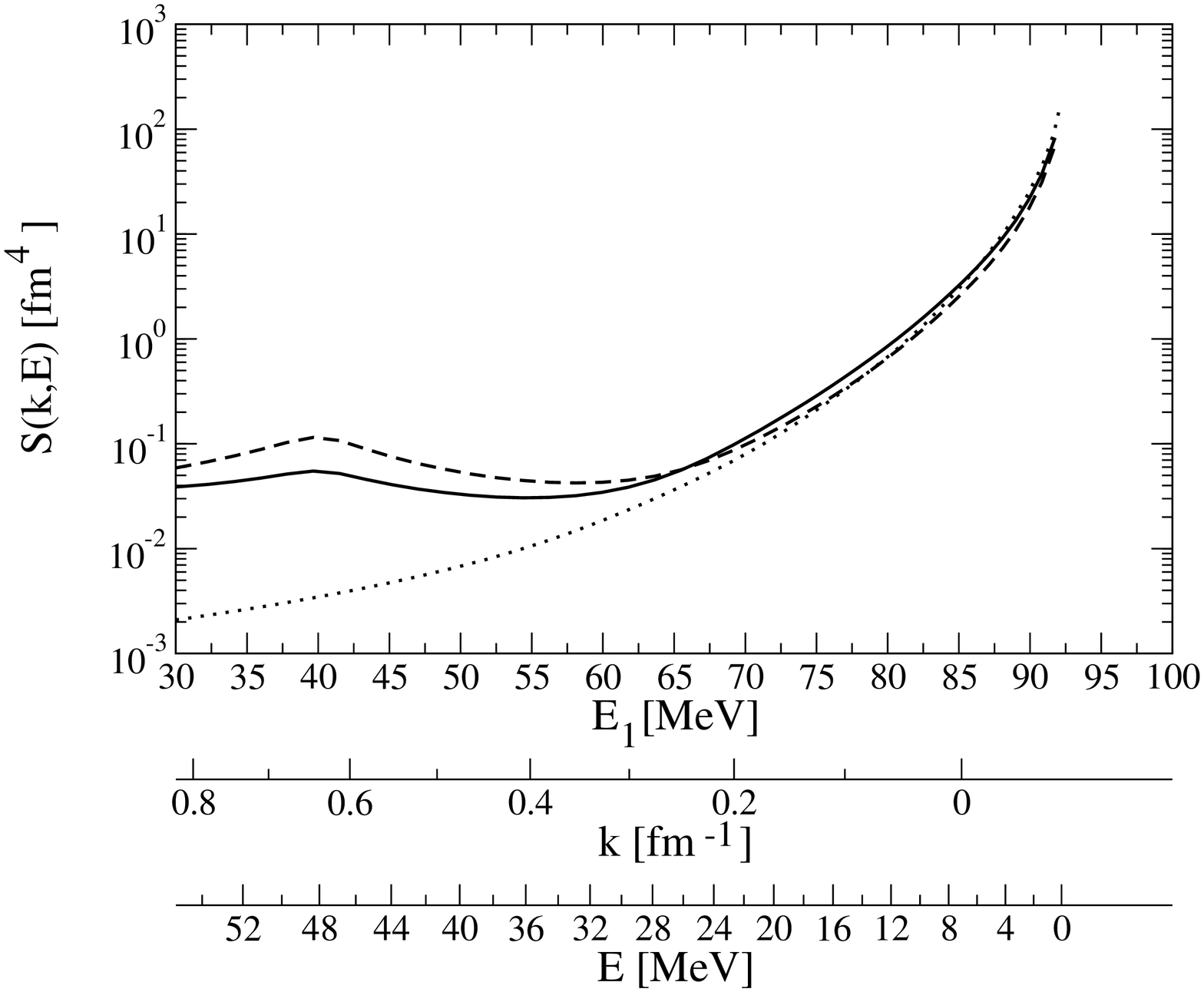,angle=-90,width=11cm}
\caption{\label{fig20}
The spectral function $S(k,E)$ for the proton knockout (dotted line),
Full results based on the form given in  Eq.~(\ref{eq9}) 
for the response functions $R_L$ (dashed line)
and Full results for the response functions $R_T$ (solid line)
for a fixed ($Q-\omega$) pair:  $\omega$= 100 MeV, $Q$= 400 MeV/c
as a function of the ejected proton energy $E_1$ for the parallel 
kinematics ${\vec p}_1 \parallel {\vec Q}$.
The corresponding values of $k$ and $E$ are also indicated.
      }
\end{center}
\end{figure}

\begin{figure}[htb]
\begin{center}
\epsfig{file=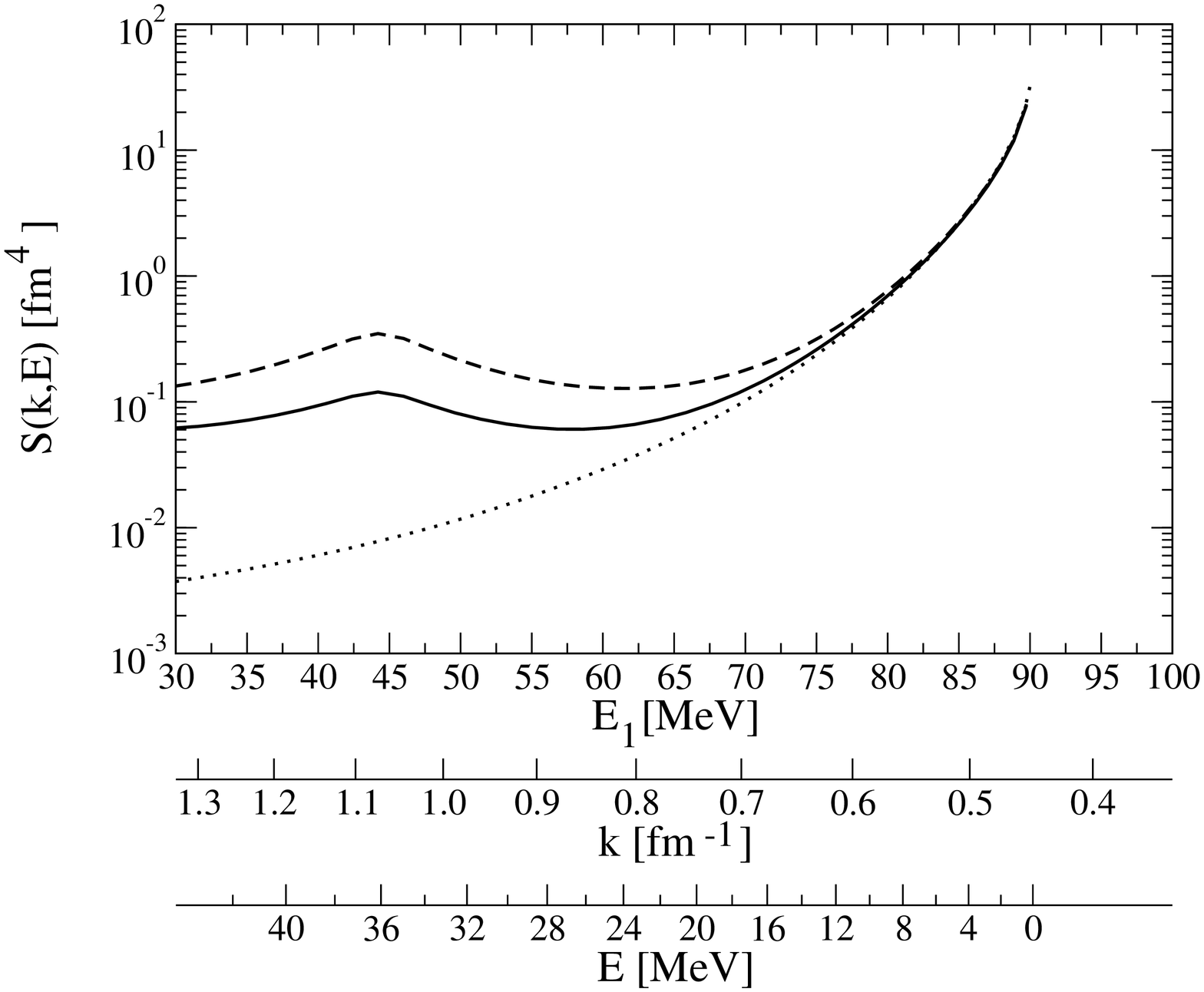,angle=-90,width=11cm}
\caption{\label{fig21}
The same as in Fig.~\ref{fig20} 
for $\omega$= 100 MeV and $Q$= 500 MeV/c.
      }
\end{center}
\end{figure}

\begin{figure}[htb]
\begin{center}
\epsfig{file=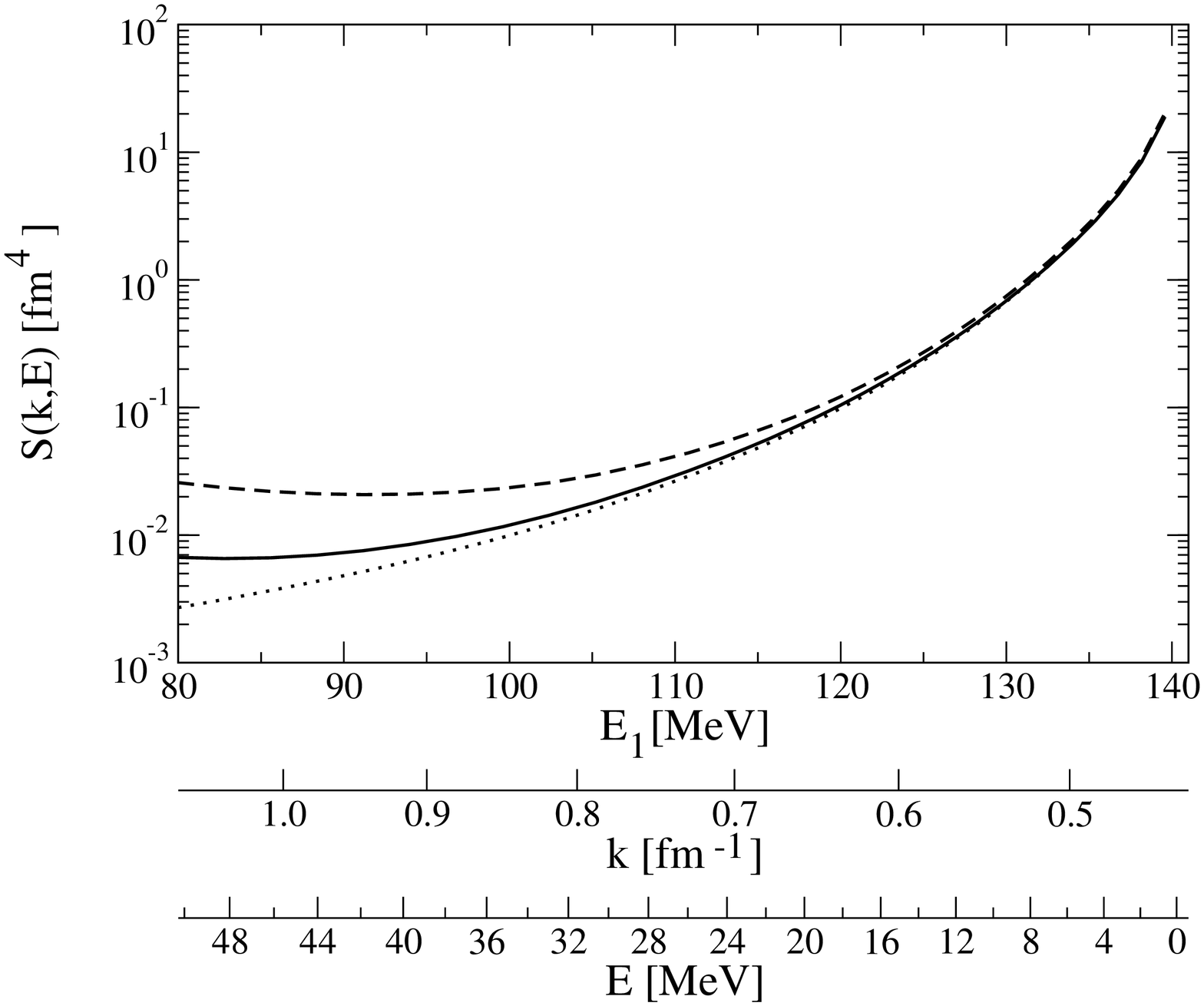,angle=-90,width=12cm}
\caption{\label{fig22}
The same as in Fig.~\ref{fig20} 
for $\omega$= 150 MeV and $Q$= 600 MeV/c.
      }
\end{center}
\end{figure}

In all these cases 
$k$ and $E$ get very small
when $E_1$ approaches its maximal value.
This is illustrated in Figs.~\ref{fig20}--\ref{fig22}.
We restrict ourselves to the upper
end of the energy $E_1$ since only there $S$ and $S(Full)$ approach each other.

We see a very nice coincidence of $S$ 
with the full results at the upper end of $E_1$,
both for $R_L$ and $R_T$.
Thus the full cross section can be rather well represented 
by the spectral function
approximation.
%(Ich schlage vor, nur FSI23 und S fuer $R_L$ und $R_T$ zu zeigen in jeweils einem Bild).

\begin{figure}[htb]
\begin{center}
\epsfig{file=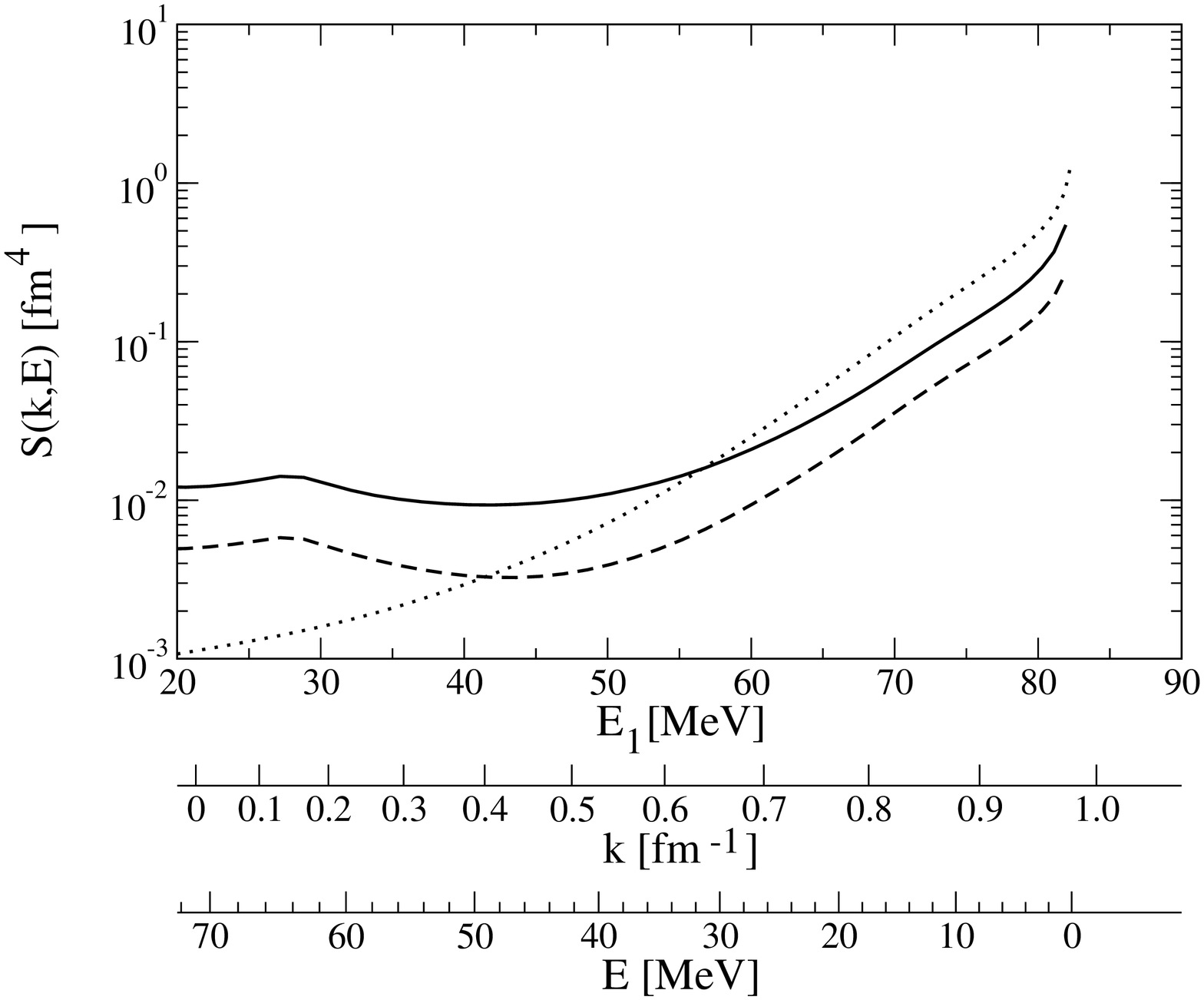,angle=-90,width=12cm}
\caption{\label{fig23}
The same as in Fig.~\ref{fig20} 
for $\omega$= 100 MeV and $Q$= 200 MeV/c.
      }
\end{center}
\end{figure}

\begin{figure}[htb]
\begin{center}
\epsfig{file=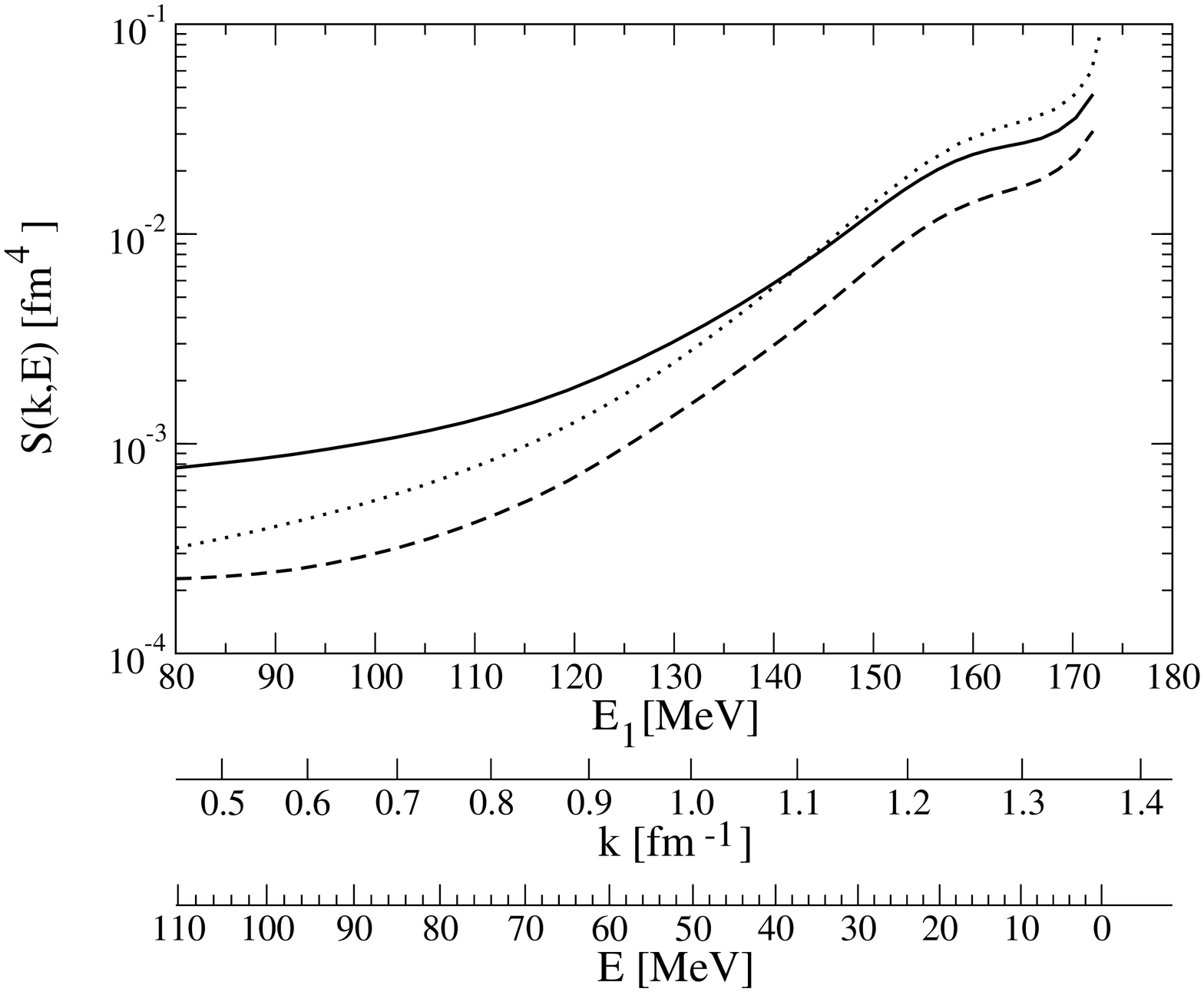,angle=-90,width=12cm}
\caption{\label{fig24}
The same as in Fig.~\ref{fig20} 
for $\omega$= 200 MeV and $Q$= 300 MeV/c.
      }
\end{center}
\end{figure}

As counterexamples one can choose: 
$\omega = 100$ MeV, $Q= 200$ MeV/c; 
%$\omega = 100$ MeV, $Q= 600$ MeV/c; 
%$\omega = 150$ MeV, $Q= 200$ MeV/c; 
$\omega = 200$ MeV, $Q= 300$ MeV/c 
shown in Figs.~\ref{fig23}--\ref{fig24}.
%Vielleicht sollten wir (100,600) und( 200,300) weglassen und nur
%zwei Gegenbeispiele zeigen.
We see indeed, that there is no agreement of $S$ 
with the full results, neither in
relation to $R_L$ nor to $R_T$ and the 
approximation using $S$ is not acceptable.

\begin{figure}[htb]
\begin{center}
\epsfig{file=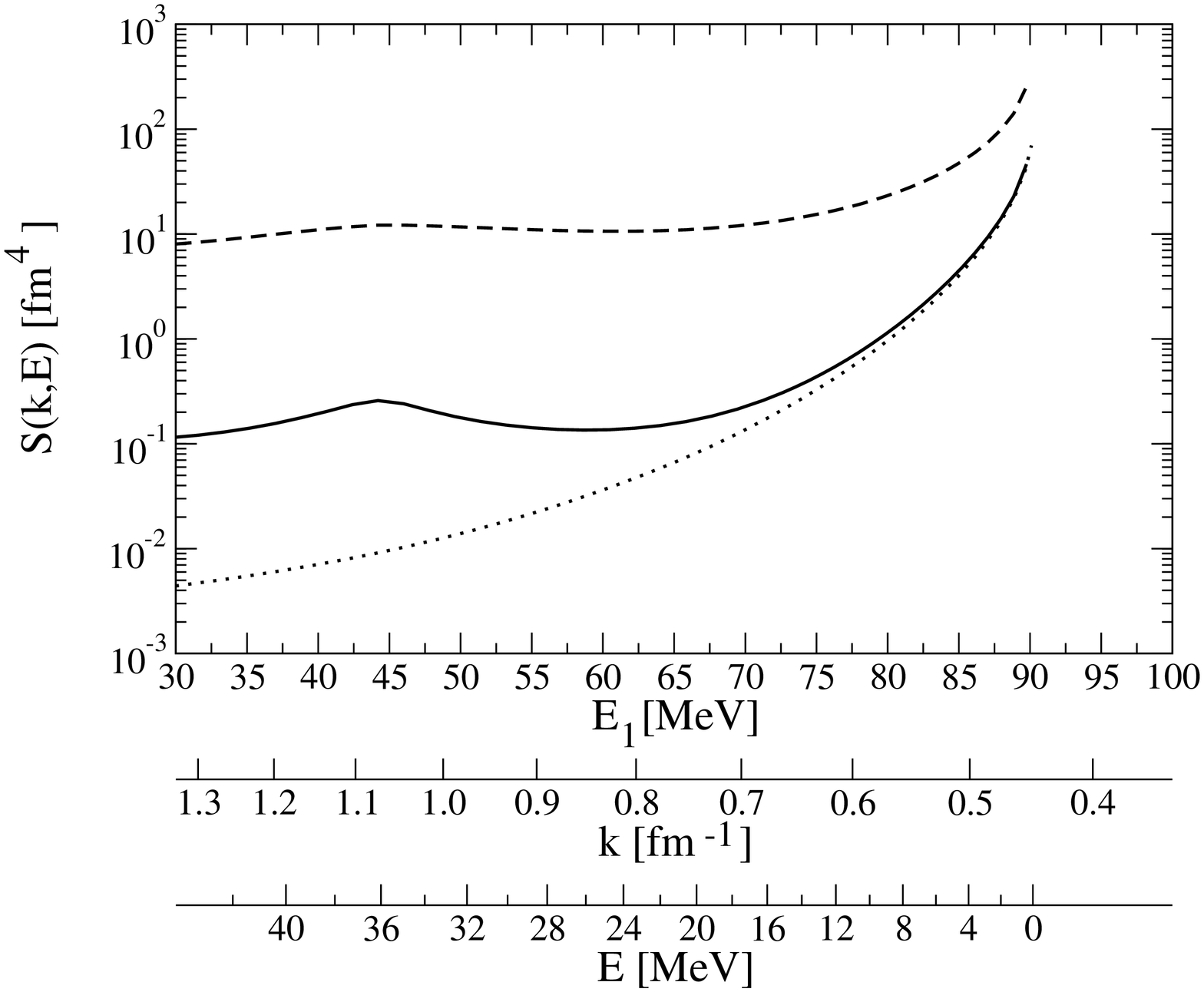,angle=-90,width=12cm}
\caption{\label{fig25}
The same as in Fig.~\ref{fig21} for the neutron knockout.
      }
\end{center}
\end{figure}

\begin{figure}[htb]
\begin{center}
\epsfig{file=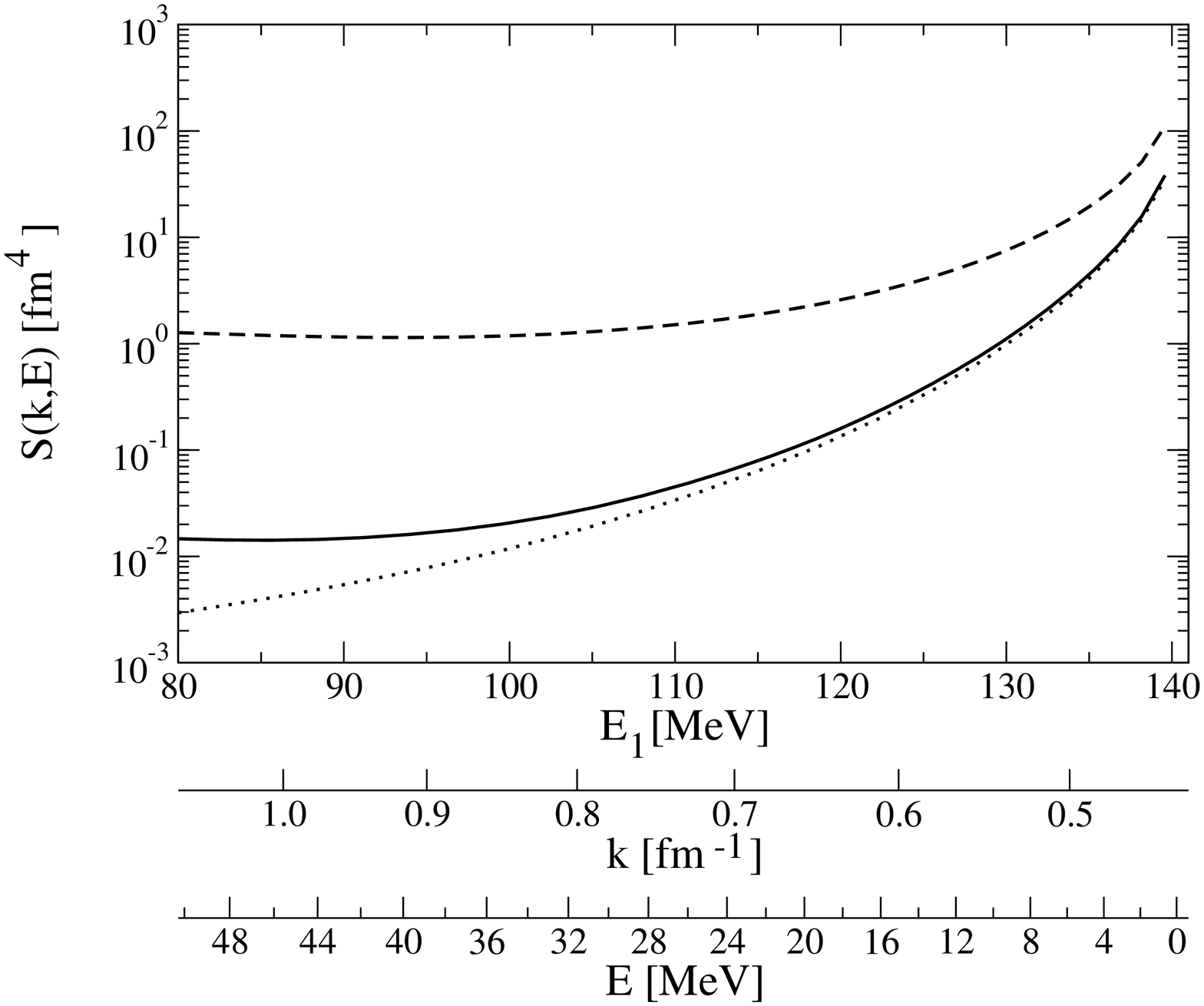,angle=-90,width=12cm}
\caption{\label{fig26}
The same as in Fig.~\ref{fig22} for the neutron knockout.
      }
\end{center}
\end{figure}

In the case of neutron knockout only $R_T$ can be approximated 
by the spectral function and
therefore the approximation of the full cross section is not suitable. 
We show in Figs.~\ref{fig25}--\ref{fig26} only two examples, 
for $\omega$= 100 MeV, $Q$= 500 MeV/c
and $\omega$= 150 MeV, $Q$= 600 MeV/c.

%say ............( Whatever you like, 
%og course out of the group we have chosen for
%the proton knock out) .
%This is shown in
%Figs.~\ref{fig31}--\ref{fig37} for the same $\omega-Q$ pairs as in Figs.~\ref{fig20}--\ref{fig26}

\begin{figure}[htb]
\begin{center}
\epsfig{file=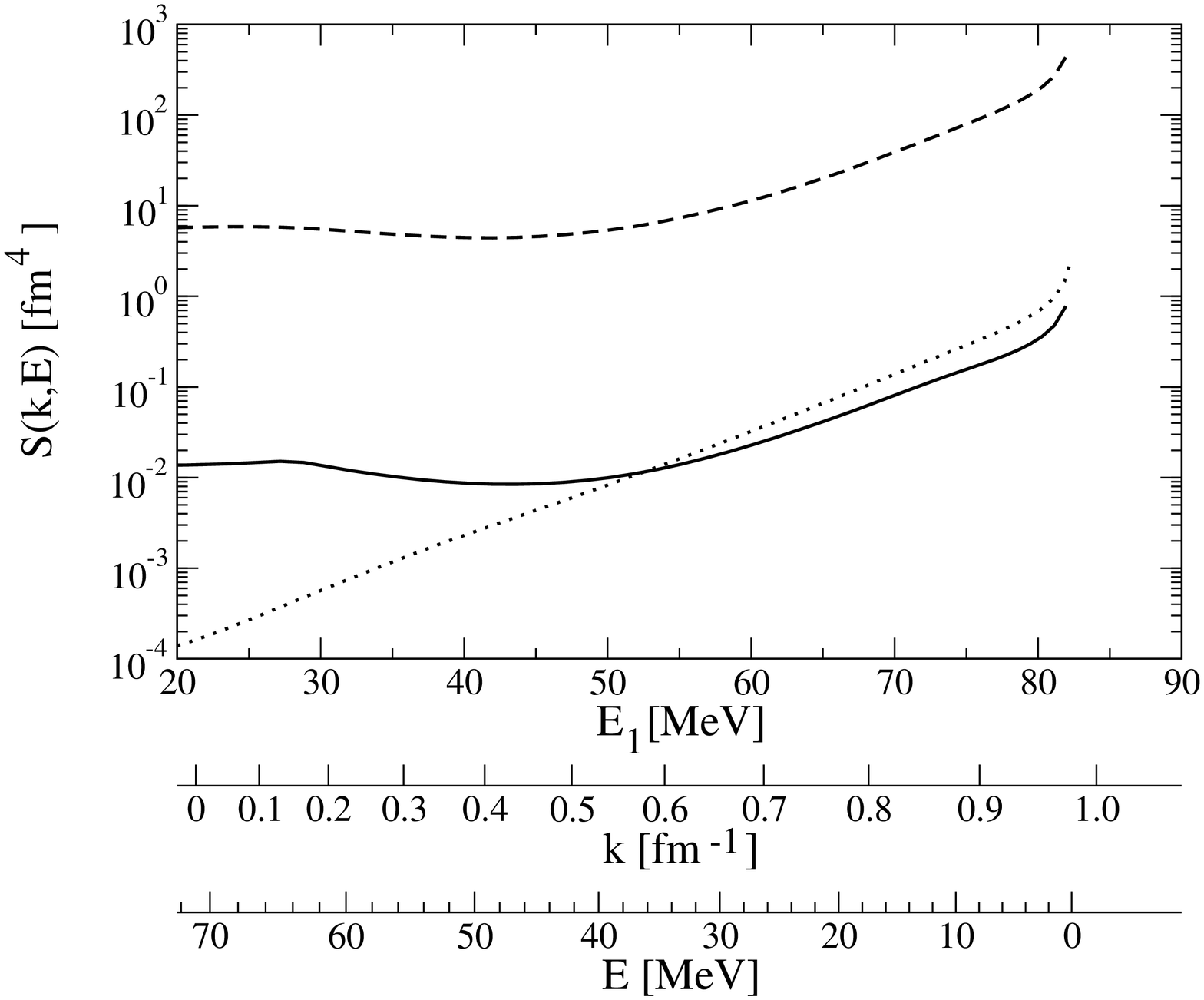,angle=-90,width=12cm}
\caption{\label{fig27}
The same as in Fig.~\ref{fig26} for the neutron knockout
and $\omega$= 100 MeV, $Q$= 200 MeV/c.
      }
\end{center}
\end{figure}

Finally one example is displayed 
in Fig.~\ref{fig27} for $\omega$= 100 MeV and $Q$= 200 MeV/c,
where the spectral function even for $R_T$ is not a sensible approximation.

%in Figs.~\ref{fig38}--\ref{fig41}, where especially $S_L(Full)$ is far away from $S$. 
One has to conclude that the
spectral function $S$ is not a good tool to analyze neutron knockout inside the
domain $D$ except for special $Q-\omega$ pairs at the upper end of $E_1$ in
case of the transversal
response.

%Kann man dabei nicht den magnetischen Formfaktor des Neutrons extrahieren?
%Wenn man eine Seperation nach $R_L$ und $R_T$ vornehmen wuerde, 
%dann waere der  $R_T$ Anteil
% bekannt und dieser waere bis auf einen bekannten 
%Faktor gleich $S$ mal $G_M^2$. Waere dies 
%nicht ein
%brauchbarer Vorschlag ? Auf diese Art koennte  man auch die Formfaktoren 
%der Protonen bestimmen
%als Test dieser Methode. Wenn Sie dies fuer richtig halten, 
%koennte man dies noch anfuegen
%in dem Papier.

Finally we would like to add a remark on the extraction of
electromagnetic form factors of the nucleons. In the case that the FSI23
approximation is valid or in other words the use of the spectral function
is justified, the electromagnetic form factors are
directly accessible. As seen in Eq.~(11) a $L-T$ separation provides direct access
to both, $G_E$ and $G_M$. It appears interesting to check that
approach firstly in the case of the proton knockout, where the form
factors are known. In the case of the neutron knockout the transverse
response function $R_T$ can be well controlled under the kinematic conditions
discussed above and therefore access to $G_M^n$ appears possible. In the case of
$G_E^n$ it might also work at higher energy and momentum transfers,
which are however outside the kinematic regime investigated in this study.

\section{\label{sec4}Summary}

We reviewed briefly the formulation of the full treatment of
the final state interaction for  the process
${}^3{\rm He}(e,e'N)$ in the Faddeev scheme. 
We showed that the processes underlying the concept 
of the spectral function are just the very first
two diagrams in an infinite  series of diagrams caused by rescattering
and complete antisymmetrization. The spectral function $S $ is directly
related to both response functions, $R_L$ and $R_T$, under those simplifying
assumptions. We used the same formal relation  which leads to $S$
but now working with the  response functions which include the
complete final state interaction. This leads to quantities $S_L( Full)$
and $S_T(Full)$, which can be compared to $S$. The comparison
was restricted to a kinematical regime where a non-relativistic treatment
appears mostly justified. Thus we restricted $E_{3N}^{c.m.}$ to be below the
pion threshold, more precisely to stay below 150 MeV and the magnitude of
the photon  momentum $\vec Q $ to be below 600 MeV/c. This defined a domain
$D$ in the $Q-\omega$ plane. The kinematical conditions for parallel 
knockout lead then to a quadratic equation connecting $Q-\omega$ to $k-E$, the
missing momentum and missing energy. Thus the domain $D$ is mapped into a
domain $D^\prime$ in the $k-E$ plane and vice versa.
Our results show that for proton knock out $S_L(Full )$ and $S_T(Full)$
agree with the approximate quantity $S$ (appropriately corrected by
electromagnetic form factors and kinematical factors) if both $k$ and $E$ are
very small. Unfortunately this is not always the case and therefore
the validity of that  approximation has better to be checked in each
case. For the rest of the domain $D^\prime$ in the $k-E$ plane $S$ is not a valid
approximation. Specifically there occur intriguing cases, where inside
$D^\prime$ $S$ coincides with the most simple approximate treatment of
the process, namely pure PWIA. This suggest a direct view into the
3He wave function. However, this is quite misleading under the kinematics
investigated  here since even the complete antisymmetrization totally
destroys that simple picture  not to speak of the final state
interaction of the knocked out nucleon with the other two.

In the case of neutron knockout, only $R_T$ can be approximated by $S$ under
certain kinematical conditions (low $k-E$ values). In the case of $R_L$ the
smallness of $G_E^n$ in relation to $G_E^p$ leads always to an important
contribution of the absorption of the photon by the two protons, which
then by final state interaction knock out the neutron. So $R_L$
in the case of neutron knockout cannot be approximated by $S$ in
the kinematic regime investigated in our study.
Finally we would like to note that the concept of $S$ might be useful to
extract electromagnetic nucleon form factors if the kinematical
conditions are suitable.

\acknowledgments
This work was supported by the Polish Committee for Scientific Research 
under grant no. 2P03B00825, by the NATO grant no. PST.CLG.978943
and by DOE under grants nos. DE-FG03-00ER41132 and DE-FC02-01ER41187.
One of us (W.G.) would like to thank the Foundation for Polish Science 
for the financial support during his stay in Krak\'ow.
The numerical calculations have been performed on the cray SV1 and T3E
of the NIC in J\"ulich, Germany.

\end{document}